\documentclass[pdflatex,sn-mathphys-num]{sn-jnl}
\usepackage{graphicx}%
\usepackage{multirow}%
\usepackage{amsmath,amssymb,amsfonts}%
\usepackage{amsthm}%
\usepackage{mathrsfs}%
\usepackage[title]{appendix}%
\usepackage{xcolor}%
\usepackage{textcomp}%
\usepackage{manyfoot}%
\usepackage{booktabs}%
\usepackage{algorithm}%
\usepackage{algorithmicx}%
\usepackage{algpseudocode}%
\usepackage{listings}%
\usepackage{todonotes}
\usepackage{subcaption}%
\renewcommand{\figurename}{Figure}

\begin{document}

\title[Mobility Behavior Evolution During Extended Emergencies: Returners, Explorers, and the 15-Minute City]{Mobility Behavior Evolution During Extended Emergencies: Returners, Explorers, and the 15-Minute City}


\author[1]{\fnm{Omid} \sur{Armantalab}}\email{oarmantalab2@huskers.unl.edu}
\author[2]{\fnm{Jason} \sur{Hawkins}}\email{jfhawkin@ucalgary.ca}
\author*[1]{\fnm{Wissam} \sur{Kontar}}\email{wkontar2@nebraska.edu}

\affil[1]{\orgdiv{Civil and Environmental Engineering}, \orgname{University of Nebraska-Lincoln}, \country{United States}}

\affil[2]{\orgdiv{Civil Engineering}, \orgname{University of Calgary}, \country{Canada}}


\abstract{Understanding human mobility during emergencies is critical for strengthening urban resilience and guiding emergency management. This study examines transitions between \textit{returners}, who repeatedly visit a limited set of locations, and \textit{explorers}, who travel across broader destinations, over a 15-day emergency period in a densely populated metropolitan region using the YJMob100K dataset. High-resolution spatial data reveal intra-urban behavioral dynamics often masked at coarser scales. Beyond static comparisons, we analyze how mobility evolves over time, with varying emergency durations, across weekdays and weekends, and relative to neighborhood boundaries, linking the analysis to the 15-minute city framework.

Results show that at least two weeks of data are required to detect meaningful behavioral shifts. During prolonged emergencies, individuals resume visits to non-essential locations more slowly than under normal conditions. Explorers markedly reduce long distance travel, while weekends and holidays consistently exhibit returner-like, short distance patterns. Residents of low Points of Interest (POI) density neighborhoods often travel to POI rich areas, highlighting spatial disparities. Strengthening local accessibility may improve urban resilience during crises.

\textit{Full reproducibility is supported through the project website: \href{https://github.com/wissamkontar}{https://github.com/wissamkontar}}
}


\maketitle
\section{Introduction}\label{sec:introduction}

The growing frequency and severity of natural disasters present profound challenges for societies worldwide, threatening lives, disrupting economies, and straining infrastructure. Economic losses have risen sharply in recent decades, driven by both intensifying hazards and increasing exposure of human settlements \citep{yuryeva2023increasing}. This trend underscores the urgent need for more effective preparedness and response strategies.

Understanding how individuals move and behave during crises is central to effective emergency management. Mobility patterns shape rescue operations, public safety measures, and long-term recovery strategies. Prior research highlights the value of analyzing collective behavioral responses in large-scale emergencies to optimize protocols and reduce risk \citep{bagrow2011collective}, as well as the importance of predictive modeling to anticipate mass displacements and guide resource allocation \citep{song2014prediction}. Recent studies show that diverse hazards, from urban floods to pandemics, can significantly disrupt established travel patterns, requiring flexible, data-informed planning approaches \citep{tang2023resilience, li2022spatiotemporal}. Mobility also mediates how individuals encounter risk and access critical services. For example, \citet{liu2023beyond} illustrate how movement behaviors shape exposure to environmental hazards, while \citet{l2024keeping} identify limitations in emergency transport systems, including fragmented information and outdated technologies, that hinder rapid response. The COVID-19 pandemic exemplifies mobility transformations under crisis, with restrictions producing sharp behavioral shifts, particularly in dense urban areas \citep{santana2022changes, santana2023covid}, and disproportionately affecting disadvantaged regions \citep{bonaccorsi2020economic}.

Despite extensive research on human mobility during emergencies, it remains unclear whether well established behavioral classifications, specifically the dichotomy between returners (individuals who frequent a limited set of familiar locations) and explorers (those with broader spatial patterns), persist under disruptive conditions. This distinction has important implications. During pandemics, explorers can accelerate disease spread, while returners may amplify local risks by concentrating within neighborhoods. Understanding how these groups behave enables planners and emergency responders to design targeted interventions and allocate resources more effectively. Equally important is whether individuals maintain or shift their mobility type over prolonged emergencies, as such dynamics shape the timing and design of policies.

To date, only one recent study has investigated this dichotomy in an emergency setting. Using mobile phone data from Hurricane Ian, \citet{he2024returners} analyzed returners and explorer under crisis conditions. While their study provided valuable initial evidence that mobility behavior diverges during disasters, it has several limitations that restrict the generalizability and depth of the findings. Their observation period was limited to four days which is insufficient for classifying returners and explorers. Variations across weekdays, weekends, and holidays were not considered, nor was the evolution of behavior over time. The study also focused on low density regions, leaving urban dynamics largely unexplored. Finally, although four mobility transition groups were identified, their behavioral characteristics were not analyzed beyond simple proportions. Motivated by these gaps, this study addresses the limitations and extends the prior work by making the following major contributions:

\begin{itemize}
    \item \textbf{Extended Emergency Coverage:} We analyze a 15-day emergency period covering two full weeks, enabling an in-depth examination of mobility behavior during extended emergencies. It captures temporal variations across weekdays, weekends, and holidays, which are essential for understanding mobility patterns under crisis conditions.
    \item \textbf{Temporal Evolution of Behavior:} We investigate how mobility behavior evolves over time in both emergency and normal periods, offering insights into how individuals adapt their travel patterns. 
    \item \textbf{Urban Scale Focus:} Unlike prior research that focused on low-density non-urban regions, this study examines mobility in a densely populated urban area, providing critical insights into intra-urban dynamics.
    \item \textbf{High Resolution Spatial Analysis:} We employ a fine-grained spatial resolution of 500-meter by 500-meter grid cells to reveal localized movement patterns that might be masked at coarser scales.
    \item \textbf{Comprehensive Returner-Explorer Group Analysis:} We analyze the behavior of all four mobility transition groups, returners-returners, returners-explorers, explorers-explorers, and explorers-returners in detail by comparing how each group’s patterns differ between normal and emergency periods and how their behavior evolves over time.
    \item \textbf{Mobility Beyond Nearby Neighborhoods:} We examine the mobility behavior outside individuals’ nearby neighborhoods, a concept closely related to the “15-minute city” \cite{abbiasov202415}. We evaluate how different groups travel and spend time outside their nearby neighborhoods during both normal and emergency periods across weekdays and weekends.   
\end{itemize}

\section{Results}\label{sec:results}

This study uses the YJMob100K dataset, an open-source, anonymized collection of human mobility trajectories from Yahoo Japan Corporation collected in 2023 \cite{yabe2024yjmob100k}. For this study, we specifically focused on the 15-day emergency period, comparing it with a corresponding 15-day normal period. This allows us to perform a comparative analysis between the mobility patterns during the emergency and the normal periods, thus providing insights into the shifts in mobility behavior due to the emergency situation. We analyze here the results across various scales and objectives explored in this study. We further refer the readers to the Supplementary Information (SI) document that provides further details on the analysis and complementary results. 

\subsection{Mobility Dynamics: Normal and Emergency}

The radius of gyration ($r_g$) is used in this study to quantify the spatial extent of each individual's mobility \cite{bagrow2012mesoscopic}. The results of fitting different distributions to the $r_g$ during both normal and emergency periods are presented in detail in SI. Our analysis shows that the lognormal distribution provides the best fit across both periods. During the emergency period, the estimated lognormal parameters indicate a slight decrease in $\mu$ and an increase in $\sigma$, suggesting a broader spread and subtle shifts in mobility behavior compared to the normal period. However, this pattern observed in the full-period analysis does not consistently hold when examined on a day-by-day basis, as the trend varies across the 15-day emergency window. This contrasts with findings from a previous study on hurricane-related mobility, where the trend remained stable over a shorter, 4-day period \cite{he2024returners}. This difference could be attributed to the shorter duration of that study, which was thus unable to capture the variability observed during the extended emergency period.

\subsection{Classifying Individuals as Returners or Explorers}

To classify individual mobility patterns, we distinguish between returners and explorers based on the $k$-radius of gyration ($r_g^{(k)}$), which measures the spatial dispersion of visits to an individual's top $k$ visited locations. The detailed analysis, including distribution fits, density analysis, and parameter estimates across normal and emergency periods, is discussed in the SI.

\begin{figure}[!ht]
    \centering
    \includegraphics[width=1\linewidth]{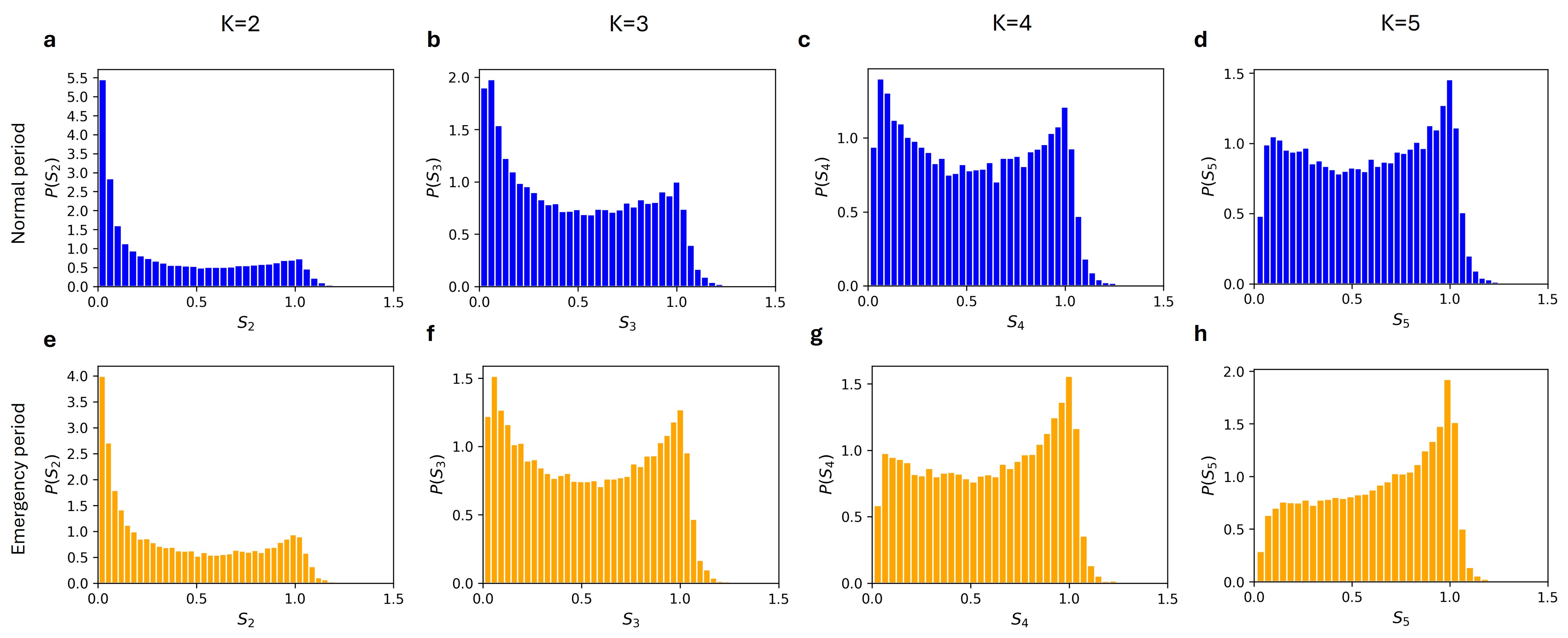}
    \caption{Distribution of the $S_k$ ratio ($r_g^{(k)} / r_g$), highlighting k-explorers ($S_k \approx 0$) and k-returners ($S_k \approx 1$) during normal (a–d) and emergency (e–h) periods.}
    \label{fig:sk_plots}
\end{figure}

Further insight into the impact of different $k$ values on mobility group identification is provided in Figure \ref{fig:sk_plots}. This figure illustrates the relationship between $r_g^{(k)}$ and $r_g$ using the metric $S_k = r_g^{(k)} / r_g$. The peaks at $S_k = 0$ and $S_k = 1$ correspond to k-explorers and k-returners, respectively. In both the normal and emergency periods, as $k$ increases, the peak of the distribution shifts from $S_k = 0$ to $S_k = 1$, indicating that individuals increasingly exhibit returner behavior when more locations are taken into account.

\begin{figure}[!ht]
    \centering
    \includegraphics[width=0.7\linewidth]{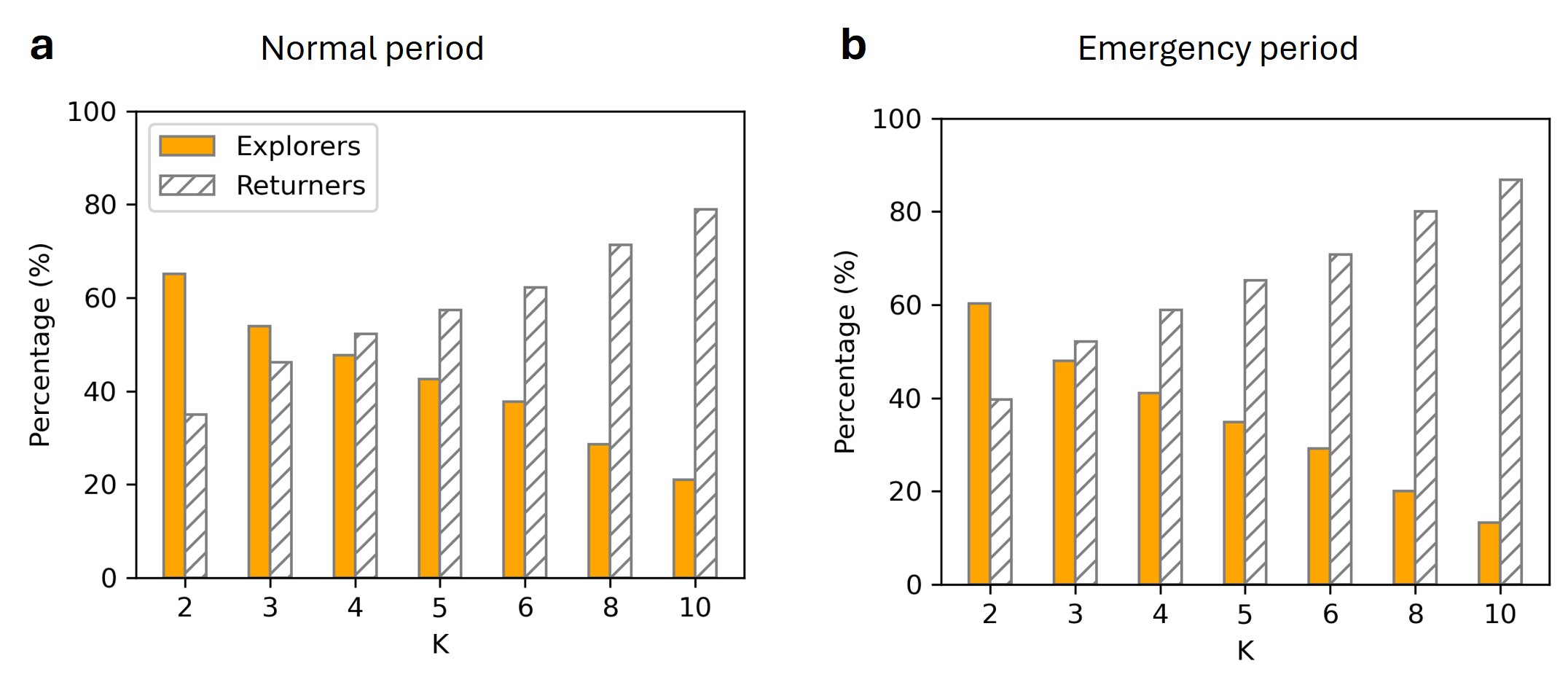}
    \caption{Proportion of k-returners and k-explorers across normal (a) and emergency (b) periods.}
    \label{fig:percentages_group}
\end{figure}

To quantify shifts in k-returners and k-explorers, residents were categorized based on both periods, and their population shares were compared. As illustrated in Figure \ref{fig:percentages_group}, the trend of an increasing share of returners and a decreasing share of explorers with higher values of $k$ remains consistent across both time periods. Specifically, in the normal period at $k=4$, the percentage of returners exceeds that of explorers, while in the emergency period, this crossover occurs at $k=3$. In the previous study examining a 4-day hurricane event in Florida \cite{he2024returners}, it was reported that the percentage of returners surpassed that of explorers at \( k = 3 \) for both emergency and normal periods. However, we argue that observing the same \( k \) value across both periods may obscure meaningful differences in mobility behavior. Specifically, the short duration of the emergency window in that study likely limited the ability to capture the broader range of activities and locations individuals typically visit during normal periods. As a result, key locations visited only under normal conditions may have been missed, leading to an underestimated \( k \) in the normal period. This highlights a limitation of using a short observation window, which may fail to capture the full extent of routine mobility patterns.

\begin{figure}[!ht]
    \centering
    \includegraphics[width=0.7\linewidth]{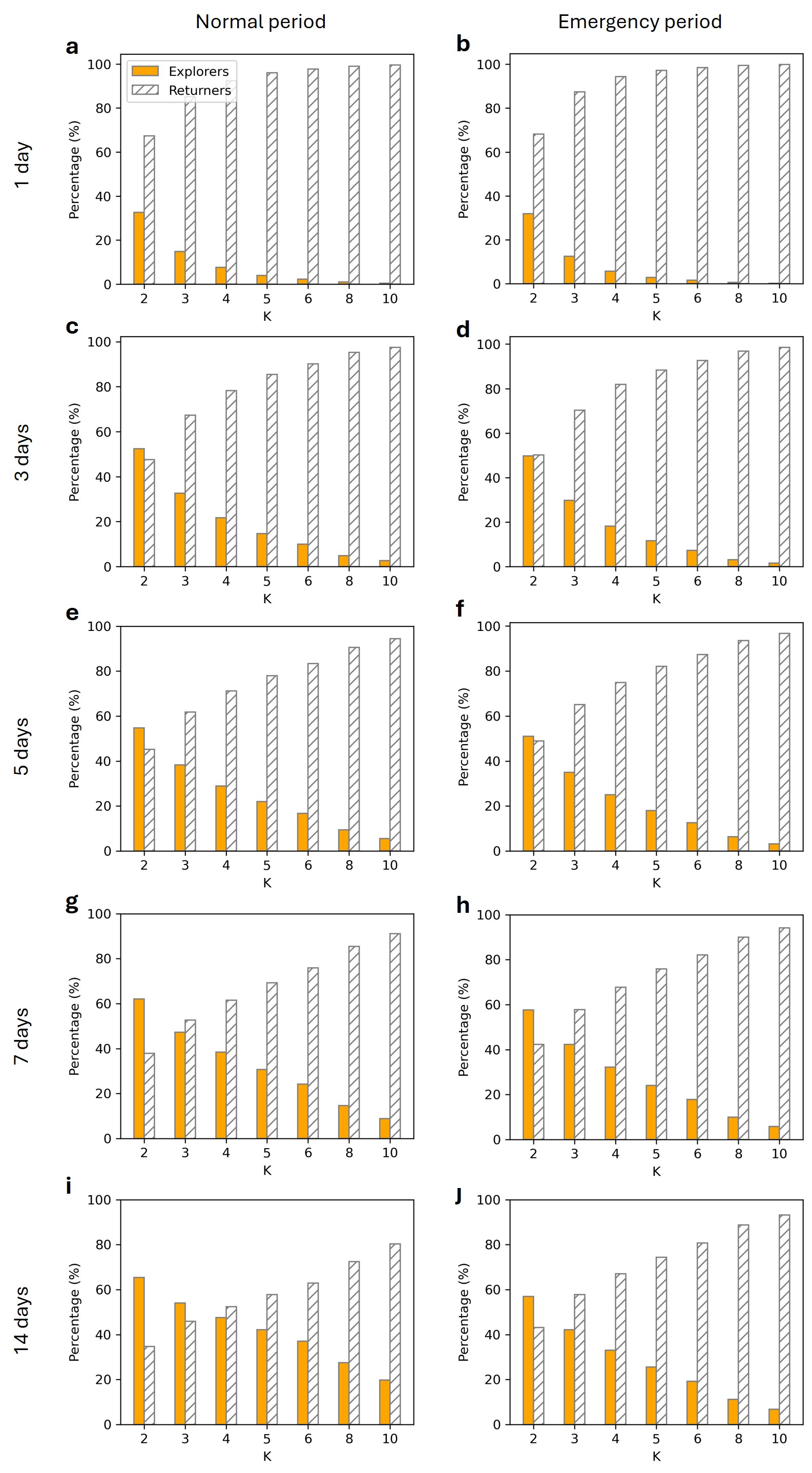}
    \caption{Percentage of k-returners and k-explorers over the first 1, 3, 5, 7, and 14 days of the normal (a, c, e, g, i) and emergency (b, d, f, h, j) periods.}
    \label{fig:1-3-5days}
\end{figure}

To investigate this further, we utilized varying time spans from our dataset to identify when differences between the emergency and normal periods become evident in terms of $k$. In Figure \ref{fig:1-3-5days}, we begin by comparing the first day of each period, then extend the comparison to the first three days, five days, seven days, and finally 14 days. This approach allows us to observe how large the dataset should be to discern differences between the emergency and normal periods. As shown, the distinction between the two periods becomes evident in the 14-day analysis (Figure~\ref{fig:1-3-5days} i, j), where in the normal period the crossover between returners and explorers occurs at $k=4$, while in the emergency period it occurs earlier, at $k=3$.

From these observations, we conclude that at least 14 days (or two weeks) of emergency data is required to effectively observe and analyze the differences between the normal and emergency periods. In this study, our dataset includes 15-day periods for both normal and emergency phases, which provides a sufficient duration to clearly examine and compare the behavior patterns between the two periods. Shorter durations tend to capture only the most essential locations which are similar across both normal and emergency periods, thereby masking less frequent destinations that appear in the normal period but not during emergencies.

To ensure that our conclusion, that at least two weeks of data are needed to observe meaningful differences, is not sensitive to the specific 14-day segment chosen, we conducted the analysis using three additional 14-day segments from the normal period. The results consistently reflected the same patterns shown in Figure~\ref{fig:1-3-5days}, confirming the robustness of our findings. Full details of this analysis are provided in the SI.

Previously, in Figure~\ref{fig:1-3-5days}, we observed that increasing the observation window from 7 to 14 days shifted the returner–explorer crossover point from $k = 3$ to $k = 4$ in the normal period. To further explore how this trend continues with longer durations, we extended the analysis to 4, 6, and 8 weeks. The results, available in the SI, demonstrate that the crossover point in the normal period increases with time: $k = 5$ for 4 weeks, and $k = 6$ for both 6 and 8 weeks. This indicates that during extended normal periods, individuals tend to visit more unique locations, and a higher $k$ is needed for returner behavior to dominate. In contrast, in the emergency period—where the crossover point remained constant at $k = 3$ for both 7- and 14-day windows—people are slower to resume visits to non-essential or discretionary locations. Their mobility remains more restricted and cautious even as the observation window increases.

\subsection{Understanding Mobility Trends for Returners and Explorers}

To investigate the mobility behaviors of returners and explorers across both periods and how they change during the emergency period, a fixed value of \( k \) must first be selected. This enables a consistent classification of individuals into returner or explorer groups by using comparable ratios across periods.
Unlike the approach in \cite{he2024returners}, which selected a fixed \( k \) based on the emergency period and analyzed both the emergency and normal periods using that fixed \( k \), we use the normal period as the baseline and adopt the corresponding fixed \( k \) value from the normal period to classify individuals and compare returner-to-explorer ratios across both periods. We believe this is more appropriate, as it allows us to clearly observe the behavioral changes that emerge during emergencies relative to typical conditions. In this study, we select \( k = 4 \) for classification, as this choice is supported by the dual peaks observed in the normal period’s $P(S_k)$ distribution at $S_k=0$ and $S_k=1$ (see Figure \ref{fig:sk_plots}), as well as the distinct crossover point in the proportions of returners and explorers at $k=4$ (Figure \ref{fig:percentages_group}). In the following sections, we examine how mobility behavior changes across different mobility groups and periods, specifically in terms of maximum distance from home \cite{alessandretti2020scales} and non-home dwelling time \cite{huang2020time}.

\subsubsection{Maximum distance from home}

\begin{figure}[!ht]
    \centering
    \includegraphics[width=0.8\linewidth]{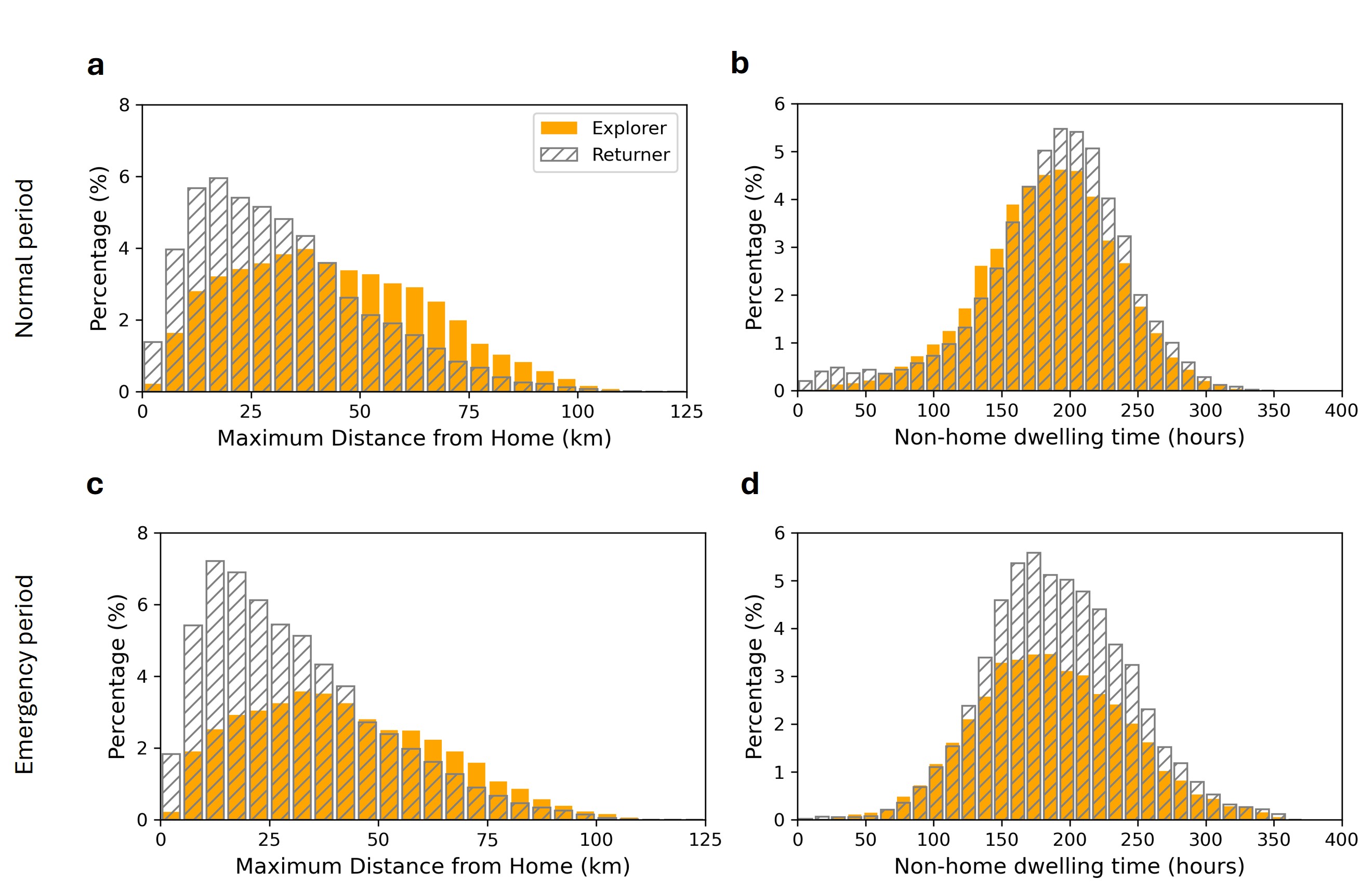}
    \caption{Maximum distance from home (a, c) and non-home dwelling 
    time (b, d) during the normal and emergency periods. The vertical axis represents the relative share of the total population.}
    \label{fig:dwel_plots}
\end{figure}

Figure \ref{fig:dwel_plots} (a and c) compares the distribution of maximum travel distances from home for both mobility groups across the two periods. As indicated in Table \ref{tab:Mann-Whitney_table_distance}, the distributions between the two groups differ significantly according to both the Kolmogorov-Smirnov and Mann-Whitney U tests. Notably, during the emergency period, explorers show a decline in long-distance travel, a finding contrary to the results observed in Florida during a Hurricane event \cite{he2024returners}. One possible explanation for this discrepancy is the difference in geographic context: our study focuses on densely populated urban areas, whereas the Florida study primarily covered low-density, non-urban regions. Results in Figure \ref{fig:dwel_plots} for the normal period show that returners tended to make shorter trips more frequently, while explorers favored longer trips. However, the emergency period seems to have altered these patterns, with the gap between the groups narrowing for longer trips and widening for shorter trips. These findings suggest that explorers may be more affected by emergency conditions, becoming more spatially constrained than returners.

\begin{table}[!ht]
    \caption{Kolmogorov-Smirnov and Mann-Whitney U test results: comparing maximum distance from home and non-home dwelling time between returners and explorers across normal and emergency periods}\label{tab:Mann-Whitney_table_distance}
    \centering
    \footnotesize
    \begin{tabular}{l c c c c}
        \hline
        & \multicolumn{2}{c}{Mann-Whitney U test} & \multicolumn{2}{c}{Kolmogorov-Smirnov test} \\\hline
         & P-value & Significance & P-value & Significance \\
        \hline
        \textbf{Emergency} & & & & \\
        Non-home Dwelling Time & 0.0 & **  & 0.0 & **\\
        Maximum Distance from Home & 0.0 & ** & 0.0 & **\\
        \hline
        \textbf{Normal} & & & &\\
        Non-home Dwelling Time & 0.0 & ** & 0.0 & **\\
        Maximum Distance from Home & 0.0 & ** & 0.0 & **\\
        \hline
        Note: ** significant at 0.01 level.
    \end{tabular}    
\end{table}

\begin{figure}[!ht]
    \centering
    \includegraphics[width=1\linewidth]{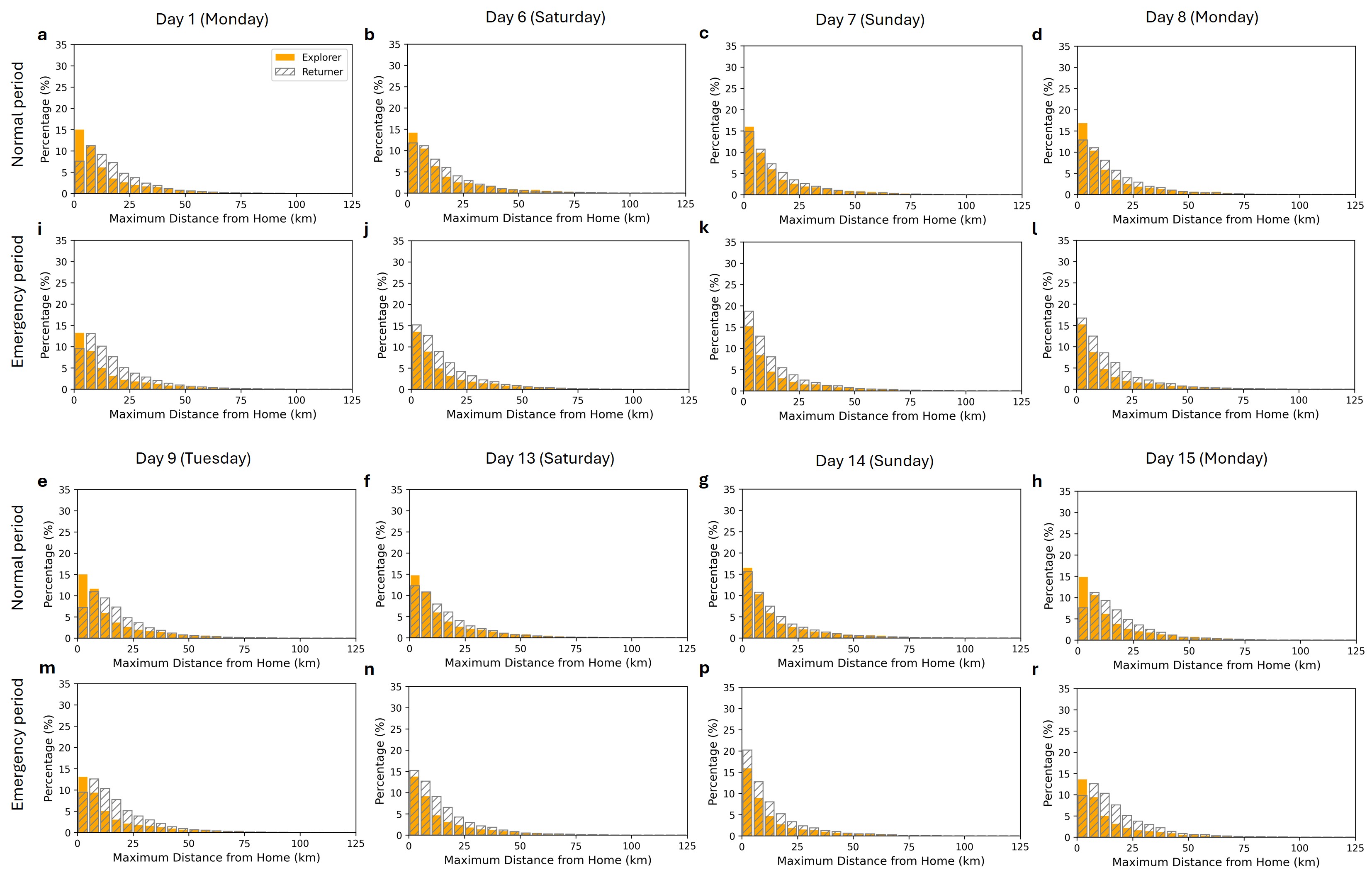}
    \caption{Maximum distance from home for each day of the normal (a–h) and the emergency periods (i–r).}
    \label{fig:distance_away_each_day}
\end{figure}

To analyze how the maximum distance from home changes over time during the emergency and normal periods, we conducted a daily analysis from day 1 to day 15, as shown in Figure~\ref{fig:distance_away_each_day}. For clarity, days with identical results are omitted; for instance, days 2–5 mirror the patterns of day 1 and are not shown. 

During the normal period, there is a slight increase in the share of explorers taking long-distance trips from home on weekends and holidays (e.g., days 6, 7, and 8) compared to weekdays. However, a notable rise in the percentage of returners making short-distance trips is observed on these days, especially on Sundays. 

A similar trend is evident during the emergency period, with Sundays showing an even more pronounced increase in short-distance trips by returners. Overall, both periods reflect a clear shift toward returner-like behavior in short-distance trips on weekends and holidays, particularly on Sundays.

\subsubsection{Non-home dwelling time}

The non-home dwelling time shows notable variation between returners and explorers, as illustrated in Figures \ref{fig:dwel_plots} (b and d). As indicated in Table \ref{tab:Mann-Whitney_table_distance}, the distributions between the two groups differ significantly. Under normal circumstances, returners compared to explorers exhibit a greater tendency to either spend very little time away from home or to be away for the majority of the time period. During the emergency period, however, explorers show more pronounced changes than returners, particularly a decrease in the time spent away from home for extended durations. This suggests that the emergency period has had a greater impact on explorers, leading them to reduce their long periods spent outside the home.

\begin{figure}[!ht]
    \centering
    \includegraphics[width=1\linewidth]{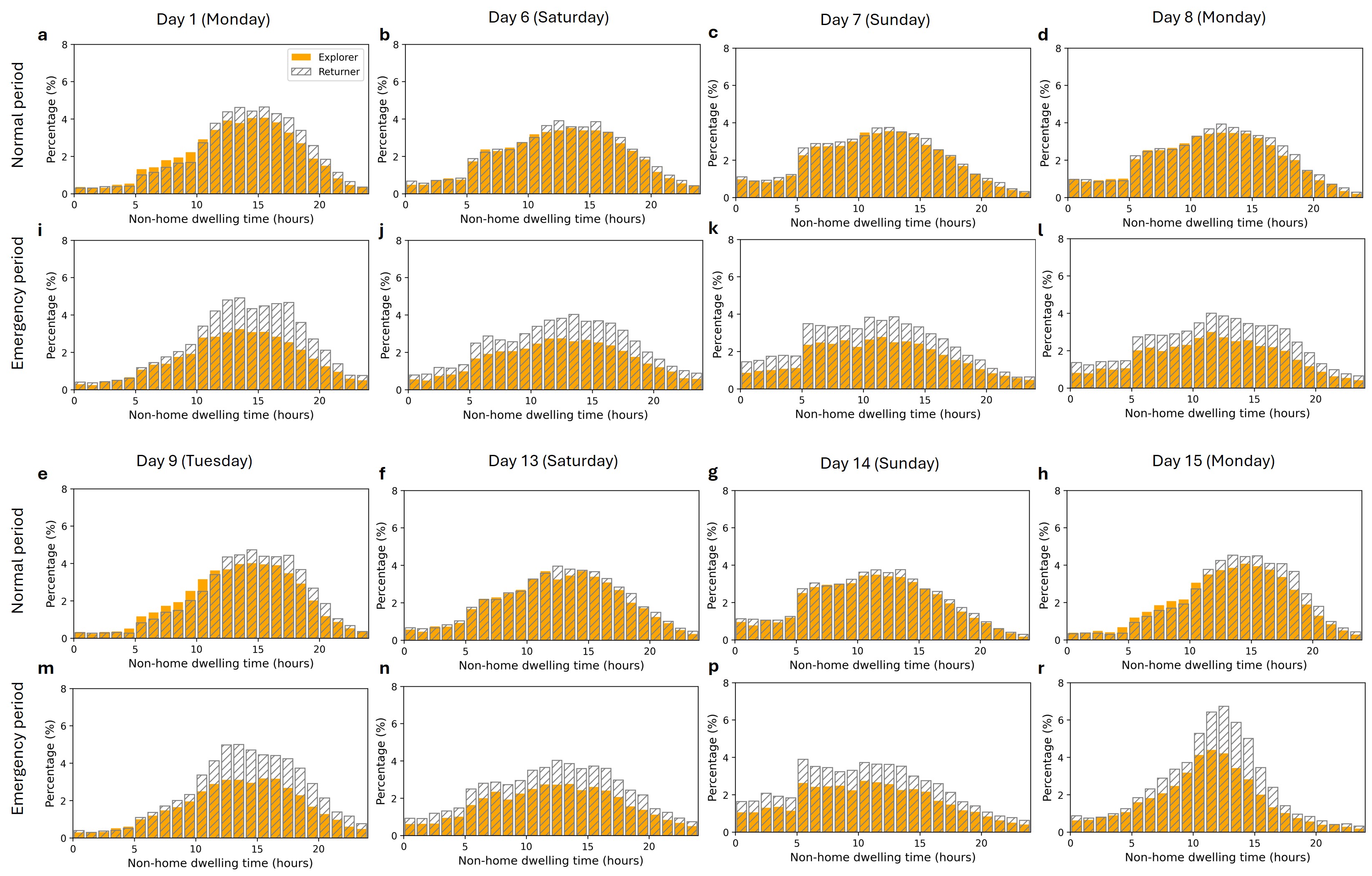}
    \caption{Non-home dwelling for each day of the normal (a–h) and the emergency periods (i–r).}
    \label{fig:time_away_each_day}
\end{figure}

We now examine how non-home dwelling time evolves over time, as shown in Figure~\ref{fig:time_away_each_day}. During the normal period, both returners and explorers show a shift toward shorter non-home durations on holidays and weekends. A similar pattern appears during the emergency period, but with an overall higher proportion of returner-like behavior across all days.

Notably, during the emergency period, there is a significant increase in the percentage of returners with shorter non-home dwelling durations on weekends and holidays, particularly Sundays, compared to weekdays. Furthermore, on the final day (day 15), both groups exhibit a noticeable shift toward shorter non-home durations and fewer long ones compared to earlier weekdays, suggesting a behavioral adjustment as the emergency period progresses.

\subsubsection{Behavioral transitions}

\begin{figure}[!ht]
    \centering
    \includegraphics[width=0.5\linewidth]{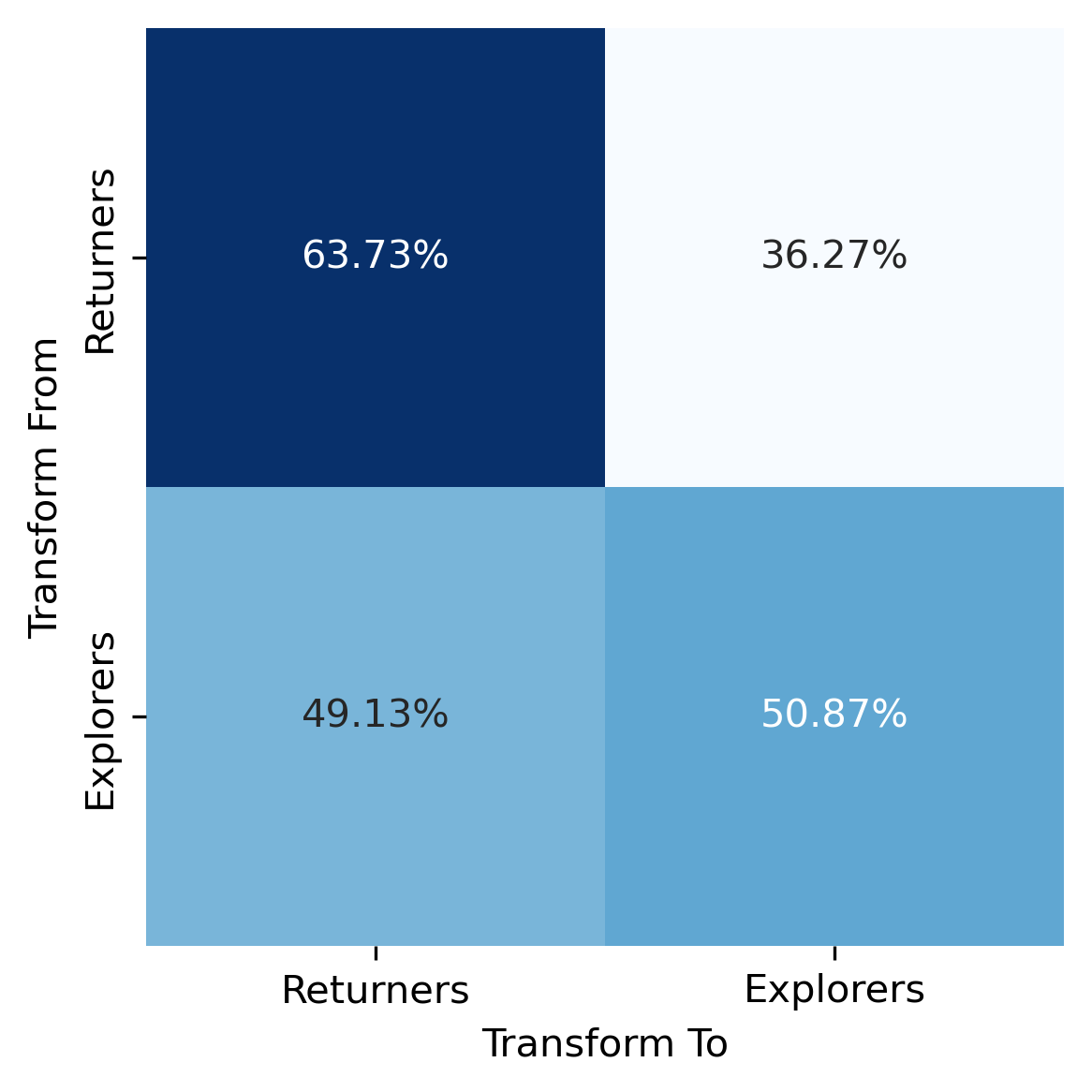}
    \caption{Transformations between returners and explorers from the normal to emergency periods.}
    \label{fig:transform}
\end{figure}

We now examine how individuals transitioned between mobility behavior types during the emergency period. As shown in Figure \ref{fig:transform}, 36.27\% of returners shifted to explorer behavior, while a larger share, 49.13\%, of explorers shifted to returner behavior. This asymmetry suggests that the emergency conditions had a stronger impact on explorers, prompting more returner-like behavior.

Having identified these four groups (returners who remained returners during the emergency period, returners who became explorers, explorers who stayed explorers, and explorers who turned into returners), this study also analyzes each groups behavior in terms of the maximum distance from home and non-home dwelling time during the emergency period day by day for all 15 days, which is discussed in detail in SI. Results revealed that returners who remained returners during the emergency period exhibited a clear behavioral shift on weekends and holidays, particularly Sundays, characterized by a substantial increase in short-distance trips and a corresponding decrease in longer trips from home. In terms of non-home dwelling time, all four groups showed a reduction in long-duration activities and an increase in short-duration ones on weekends and holidays, with the shift most pronounced among the returner–returner group. These patterns suggest a general preference for shorter, localized activities on holidays, especially among those maintaining returner-like behavior throughout the emergency.

\subsubsection{Entropy}

To further explore group differences, we also analyzed per capita real entropy across the two periods, with detailed results provided in the SI.  In summary, explorers consistently exhibited higher entropy than returners, indicating greater variability and less predictability in their mobility patterns.

\subsection{Spatiotemporal mobility and 15-minute city}

Spatial distributions of activity stops and the average duration of stay at those stops are detailed in the SI. In brief, returners consistently exhibit a higher concentration of stops and longer average stay times in the city center compared to explorers, across both normal and emergency periods. During the emergency period, mobility becomes more spatially dispersed, with increased activity outside the city center and a general reduction in average stay time. This reduction is more pronounced among explorers, suggesting that returners maintain more stable and localized routines, while explorers display greater sensitivity and adaptability to emergency conditions.

To further explore these spatial behaviors, we examine individuals' mobility relative to their nearby neighborhoods, defined using a grid system approximately aligned with the “15-minute city” framework \cite{abbiasov202415}. The “15-minute city” concept emphasizes that daily essentials should be accessible within a walkable distance from home, which has significant implications for sustainability \cite{allam202215,moreno2021introducing}. By promoting walkability, this concept helps reduce reliance on transportation which is the second-largest global energy consumer and the leading source of greenhouse gas emissions in the U.S. \cite{abbiasov202415}. Since the concept centers around access to essential needs, it becomes particularly important to examine mobility patterns during emergency periods and compare them with those during normal times. Accordingly, in this study, we analyze how different groups, explorers and returners, spend time and distance outside their nearby neighborhoods, which are assumed to be accessible via active transportation modes, during both normal and emergency periods. Details on the definition of nearby neighborhoods are provided in the SI.

\subsubsection{Time spent outside nearby neighborhoods}

\begin{figure}[!ht]
    \centering
    \includegraphics[width=0.7\linewidth]{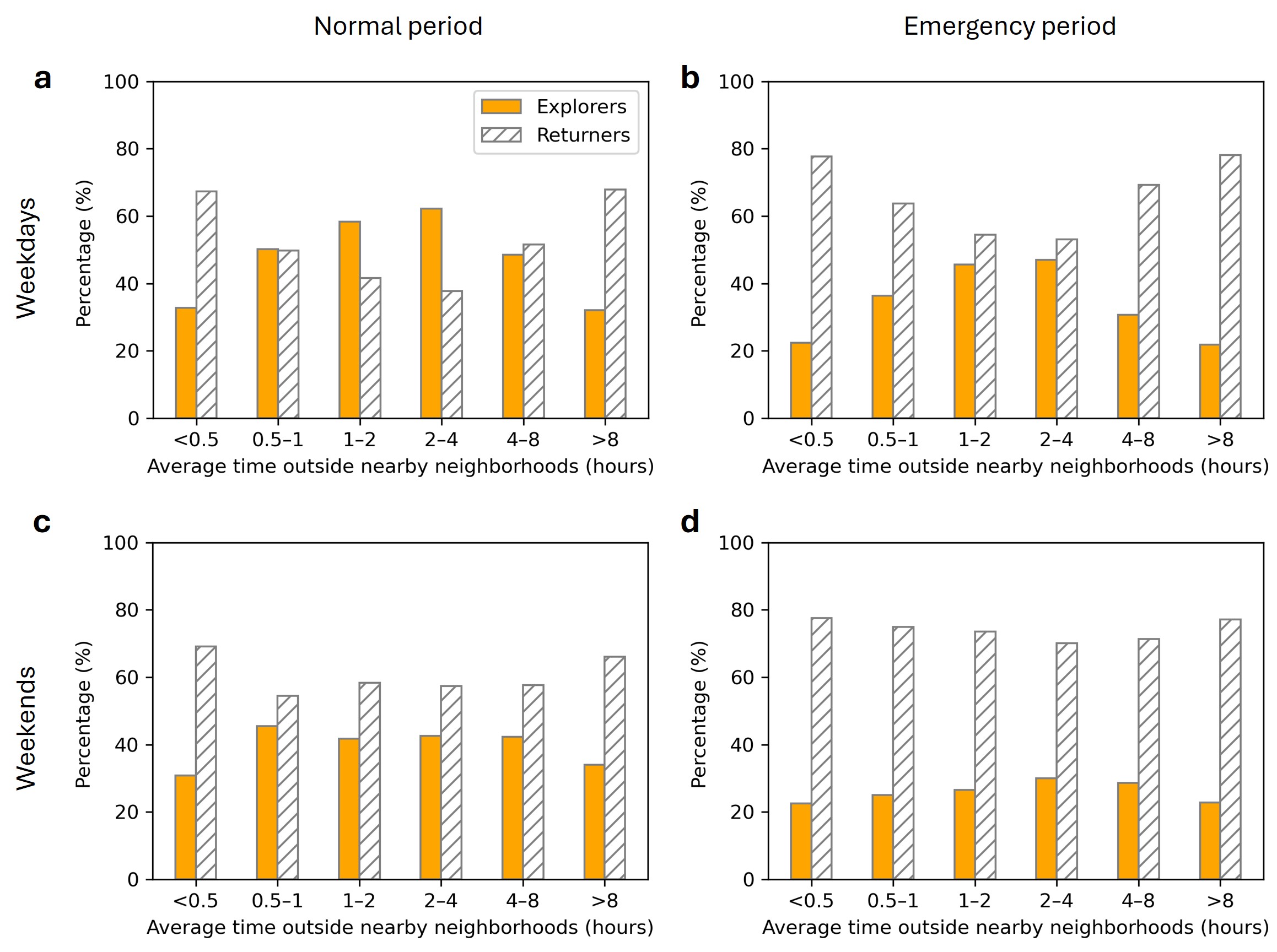}
    \caption{Distribution of returners and explorers by average daily time spent outside their nearby neighborhoods on weekdays (a, b) and weekends (c, d) during normal and emergency periods.}
    \label{fig:neighborhood_time}
\end{figure}

First, we examine how the distribution of explorers and returners varies based on the average daily time spent Outside Nearby Neighborhoods (ONN) during normal and emergency periods, separately for weekdays and weekends, as shown in Figures~\ref{fig:neighborhood_time}. Focusing on weekdays first, during the normal period (Figure~\ref{fig:neighborhood_time}a), returners account for about 63\% of users spending less than 30 minutes ONN, while explorers dominate among those spending up to 4 hours ONN. Interestingly, beyond 4 hours, the share of returners increases again. This suggests that individuals spending either very short or very long periods ONN are more likely to be returners, while moderate ONN durations are associated with a higher proportion of explorers. In the emergency period (Figure~\ref{fig:neighborhood_time}b), the share of explorers decreases across all ONN time groups, and returners outnumber explorers, indicating a general shift toward returner-like behavior.

Turning to weekends, Figure~\ref{fig:neighborhood_time}c shows that, across all ONN time groups, returners consistently outnumber explorers, reflecting a general tendency toward returner-like behavior on weekends. This pattern becomes even more pronounced during the emergency period (Figure~\ref{fig:neighborhood_time}d), where returner behavior dominates across all groups. Notably, while weekday patterns during the emergency still exhibit some variation, weekend patterns show a near-uniform dominance of returner behavior. This suggests that while weekday behavior may still reflect essential needs such as commuting, weekend mobility becomes overwhelmingly returner-like during emergencies.

\subsubsection{Distance traveled outside nearby neighborhoods}

\begin{figure}[!ht]
    \centering
    \includegraphics[width=0.7\linewidth]{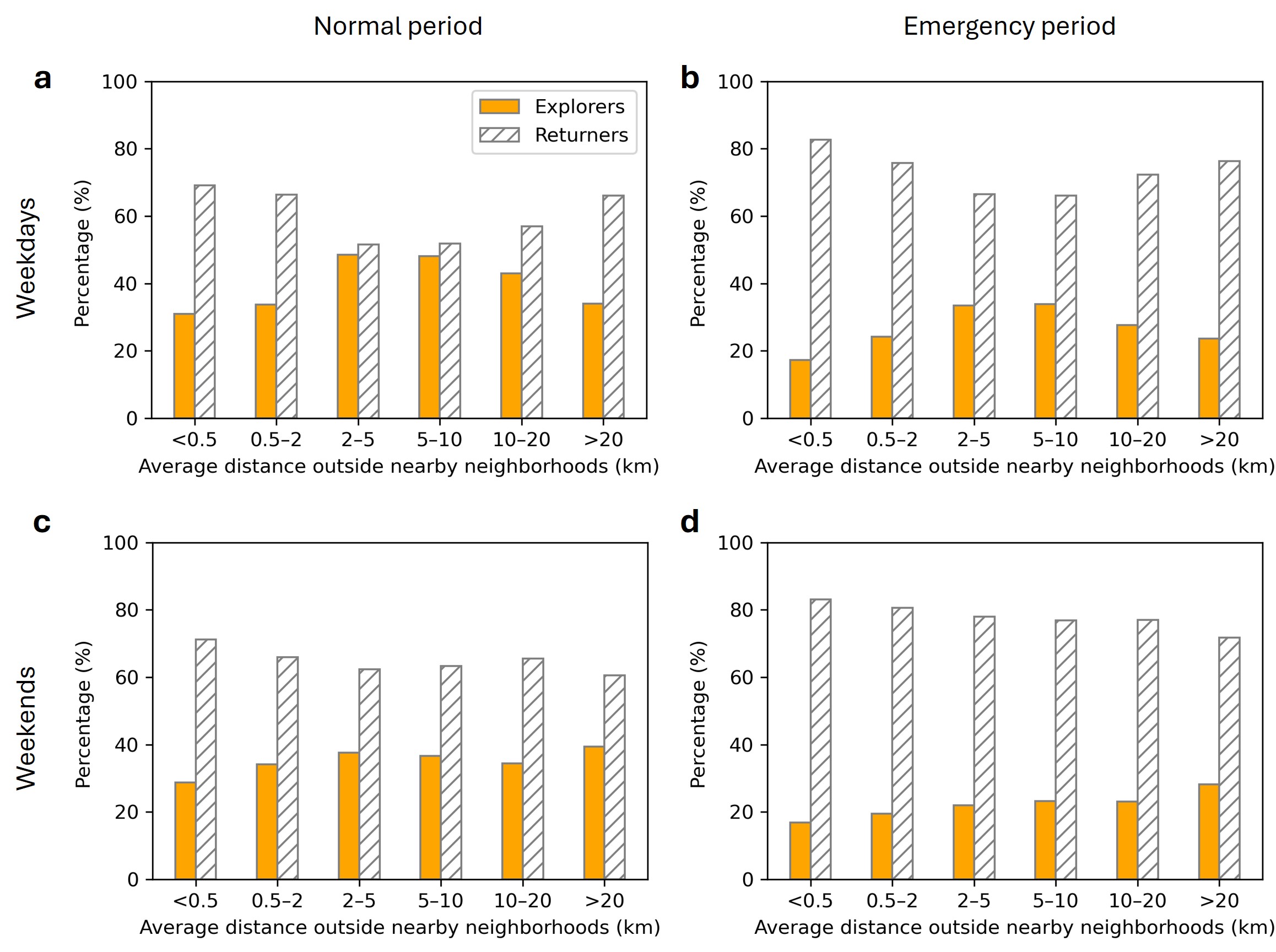}
    \caption{Distribution of returners and explorers based on average daily distance traveled outside their nearby neighborhoods on weekdays (a, b) and weekends (c, d) during normal and emergency periods.}
    \label{fig:neighborhood_distance}
\end{figure}

Now, we investigate how the composition of explorers and returners varies with the average daily distance traveled ONN. Since such travel is more likely to involve motorized modes, potentially contributing more to emissions, understanding this variation is important. With regard to weekdays, Figure~\ref{fig:neighborhood_distance}a shows that during the normal period, returners make up a larger share of users traveling either short or long distances ONN, while the proportions of explorers and returners are more balanced for moderate distances. Notably, users traveling long distances ONN—who are likely contributing more to emissions—are predominantly returners. This suggests that encouraging or enabling these individuals to access their frequently visited destinations (e.g., workplaces) within their nearby neighborhoods could lead to substantial emission reductions. 

A similar pattern appears during the emergency period, with an overall increase in the proportion of returners across all distance categories (Figure~\ref{fig:neighborhood_distance}b). This shift indicates that people are more inclined toward returner-like behavior during emergencies, reinforcing the importance of local accessibility, especially during disruptive events. On weekends in the emergency period, although returners remain the dominant group, the share of explorers gradually increases with distance traveled ONN. This suggests that some exploratory behavior persists during emergencies, especially at higher distances, though it is less prominent compared to the normal period.

\subsubsection{Spatial distribution of home locations}

\begin{figure}[ht]
    \centering
    \includegraphics[width=1\linewidth]{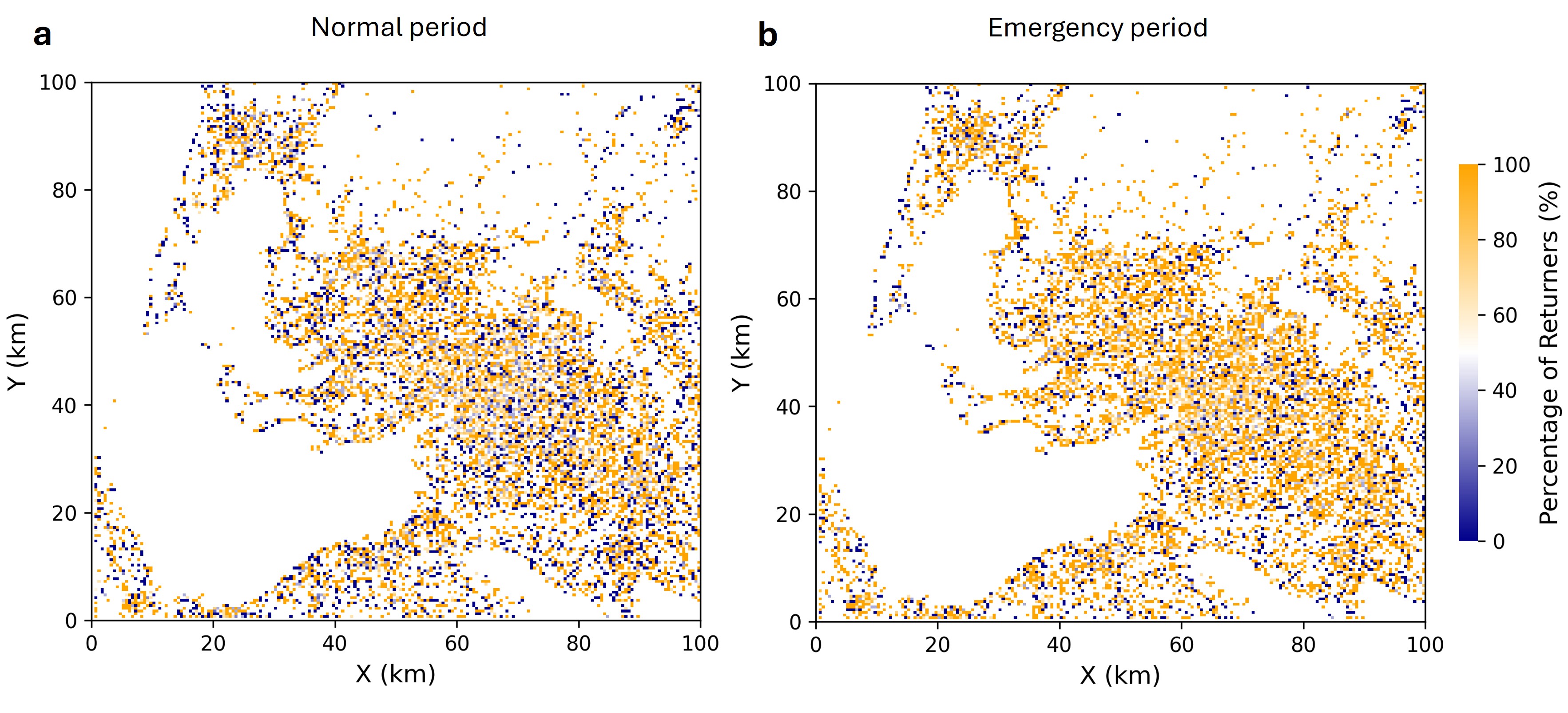}
    \caption{Percentage of returners across different areas based on individuals' home locations. Blue shades indicate areas where explorers outnumber returners, orange shades indicate areas where returners are the majority, and the white transition line (around 50\%) represents locations where the numbers of returners and explorers are approximately equal.}
    \label{fig:explorer_home_location}
\end{figure}

Figure~\ref{fig:explorer_home_location} illustrates the spatial distribution of returners and explorers based on their home locations during the normal and the emergency periods. In the normal period, a high concentration of returners is observed around coordinates $x=68$, $y=38$, corresponding to the major city center. This may be attributed to the density of POIs in this urban core, allowing residents to fulfill their daily needs locally and adopt to returner-like behaviors. In contrast, surrounding areas, especially those not too far from the central city, tend to have a greater proportion of explorers, likely due to fewer nearby amenities, requiring longer travel distances. The reverse interpretation may also hold: individuals with returner tendencies may choose to live in amenity-rich areas, while those inclined to explore may prefer more peripheral neighborhoods.

When comparing the emergency period to the normal period, there is a noticeable increase in the proportion of returners across nearly all areas. This shift is most pronounced in the central regions of the map, where the returner dominance intensifies. However, peripheral regions—likely representing rural or suburban areas—exhibit a less dramatic shift. This suggests that individuals in urban areas, where POIs are more concentrated, were more likely to reduce their travel and become returners during the emergency period, while those in less dense regions continued exploring to meet their essential needs.

\subsubsection{Number of points of interest}

Table~\ref{tab:compare_results} presents a comparison of POI-related metrics across the four behavior transition groups during the normal and emergency periods. With respect to the average number of POIs of visited locations ONN on holidays, the R-R group leads during the normal period, indicating that areas with high POI density are especially attractive to stable returners. This pattern persists in the emergency period. The same trend holds for weekdays, further supporting the idea that returners are drawn to POI-rich destinations across both time periods.

\begin{table}[!ht]
    \caption{Comparison of ONN-related metrics across behavior transitions in normal and emergency periods. \textit{R-R} refers to users who were returners in both periods, \textit{R-E} to users who changed from returner to explorer, \textit{E-E} to users who remained explorers, and \textit{E-R} to users who changed from explorer to returner. \textit{N} and \textit{E} indicate values from the normal and emergency periods, respectively.}
    \label{tab:compare_results}
    \centering
    \footnotesize
    \begin{tabular}{p{4cm} cc cc cc cc}
        \hline
        & \multicolumn{2}{c}{R-R} & \multicolumn{2}{c}{R-E} & \multicolumn{2}{c}{E-E} & \multicolumn{2}{c}{E-R} \\\hline
        & N & E & N & E & N & E & N & E \\
        \hline
        Average number of POIs of visited locations ONN on holidays  & 95.28 & 88.31 & 88.61 & 79.99 & 87.39 & 81.24 & 89.15 & 82.88 \\\\
        Average number of POIs of visited locations ONN on weekdays  & 108.03 & 102.06 & 92.19 & 84.00 & 85.64 & 79.55 & 90.30 & 85.23 \\\\
        Average number of POIs of home and nearby neighborhoods & 61.01 &  & 60.49 &  & 61.04 &  & 60.92 &  \\
        \hline
    \end{tabular}
\end{table}

Finally, regarding the average number of POIs in users' home and nearby neighborhoods, the values are significantly lower than those for visited locations. This indicates that users across all groups tend to live in areas with relatively low POI density but travel to POI-rich areas ONN. This finding highlights the importance of improving access to amenities in local neighborhoods. Encouraging users to stay within their nearby neighborhoods—thereby reducing motorized travel—may require enhancing the availability of high-POI environments locally. This strategy may be especially effective for returners, particularly those in the R-R group, who appear most responsive to POI density in destination choices.


\section{Discussion}

In this work, we endeavored to analyze the behavior of mobility under crisis through the lens of explorer-returner dichotomy. The following are some of the key findings from this study:

\begin{itemize}
    \item At least two weeks of emergency data are necessary to effectively capture mobility differences between normal and emergency periods.
    \item During extended emergencies, individuals are slower to resume visits to non-essential or discretionary locations than in normal times.
    \item During the emergency period, explorers significantly reduced long-distance travel, suggesting a general contraction in activity spaces. This trend differed from findings in less dense regions, underscoring the role of urban form in shaping mobility responses to emergency periods.
    \item In both the normal and emergency periods, weekends and holidays, especially Sundays, show that people tend to exhibit more returner-like behavior in short-distance trips from home. 
    \item Behavioral transition analysis revealed that nearly half of explorers (49.13\%) shifted to returner behavior during the emergency, compared to 36.27\% of returners who became explorers. This imbalance shows that explorers were more strongly influenced by emergency conditions.
    \item On weekdays, users traveling long distances outside nearby neighborhoods, who likely contribute more to emissions, are predominantly returners, highlighting the potential environmental benefit of improving local access to frequently visited destinations.
    \item Regardless of group, most individuals resided in neighborhoods with relatively low POI density but frequently traveled to POI-rich areas outside their nearby neighborhoods, highlighting the need to improve local amenity access and diversity to reduce long-distance travel and support the resilience and sustainability aims of the 15-minute city concept.
\end{itemize}

Together, these findings reveal how returners and explorers adjust their mobility during emergencies, how behaviors evolve as conditions persist, how patterns differ across days of the week, and where these changes emerge. Detecting such shifts through extended temporal analysis and high-resolution spatial data provides a foundation for designing more effective and context-aware interventions. Our findings suggest several directions for improving urban resilience and emergency response strategies. 

\noindent \paragraph{Extended monitoring:} The need for extended observation periods to detect behavioral changes indicates that policy monitoring systems should span several weeks, especially during prolonged emergencies, to capture when and where behavioral stabilization occurs. As observed in this study, individuals were slower to resume visits to non-essential locations under emergency conditions. Understanding the reasons behind this hesitation can inform targeted interventions to support the resumption of discretionary activities, particularly during extended emergencies.

\noindent \paragraph{Local accessibility:} The reduced long-distance travel during emergencies among explorers, along with the concentration of returners in POI-rich central areas, highlights the crucial role of local accessibility in shaping resilient mobility behavior. These patterns suggest that individuals are more likely to adopt returner-like, localized travel when essential services are nearby especially in dense urban environments. In contrast to findings from low-density regions, our results show that people in high-density settings tend to restrict their movement during emergencies, relying more on nearby amenities. The shift toward localized behavior is especially evident on weekends and holidays, particularly Sundays, when returner-like patterns become more pronounced. These findings underscore the need for equitable distribution of amenities across neighborhoods to reduce dependency on motorized travel, enhance behavioral adaptability in emergencies, and support long-term goals of sustainable and resilient urban development.

\noindent \paragraph{Temporal adaptability:} The observed divergence in behavior between weekends and weekdays calls for temporally adaptive policies. This study found that on holidays, particularly Sundays, during the emergency period, individuals, especially those consistently exhibiting returner-like behavior, tended to make more short-distance trips and engage in shorter-duration non-home activities. These natural reductions in mobility and activity time suggest that people may self-limit their behavior on certain days without the need for strict interventions. Therefore, rather than enforcing uniform restrictions across all days, emergency policies should be aligned with these temporal behavior patterns. For example, easing restrictions or adjusting service availability on weekends could accommodate essential short trips without encouraging long-distance travel, while weekday measures might more directly target work-related mobility.

\noindent \paragraph{Behavioral transitions:} The four-group classification system used in our analysis reveals that not all individuals respond to emergencies in the same way, even if they share the same baseline behavior. As such, communication strategies and interventions should be personalized where possible. For instance, individuals who switch from returner to explorer behavior during emergencies may require different forms of outreach and support than those who remain returners. Recognizing these transitions and planning accordingly could improve the effectiveness of emergency measures.

\noindent \paragraph{Urban design and planning:} The ONN framework used in this study aligns with the goals of the 15-minute city concept by demonstrating that individuals—especially returners—naturally shift toward more localized behavior during emergencies. Our results show that during the emergency period, returners became the dominant group across all categories of time spent and distance traveled outside nearby neighborhoods, particularly on weekends. However, even returners tended to travel to POI-rich areas beyond their home neighborhoods, reflecting a mismatch between residential locations and amenity access. This suggests that while people are willing to localize their mobility under pressure, the availability of nearby services remains a limiting factor. Enhancing walkability and increasing the POI density of residential neighborhoods can reduce the need for longer-distance, potentially motorized trips and support both emergency resilience and long-term sustainability. Policymakers should prioritize development patterns that bring services closer to where people live, promoting returner-like behaviors that align with both environmental and preparedness objectives.

\noindent \paragraph{Fine-grained evidence:} Our results highlight the importance of using fine-grained data and analysis to inform urban policy. County or city level analyses may obscure important intra-urban differences in behavior, leading to misdirected interventions. As demonstrated in our 500 m $\times$ 500 m grid-cell analysis, policies tailored to neighborhood-scale dynamics are more likely to resonate with actual patterns of movement and behavior. Investing in such granular data infrastructures and analytic capacity should be a priority for cities aiming to build adaptive and inclusive emergency response systems.\\

This study is subject to some limitations. First, although our analysis was conducted at a relatively fine spatial resolution of 500-meter by 500-meter grid cells, finer than those used in prior studies, more precise data at even higher spatial granularity could yield a more accurate understanding of localized behavioral responses. Second, the dataset used in this study captures location information at 30-minute intervals. While sufficient for identifying general mobility patterns, data collected at shorter time intervals could provide a more detailed picture of movement dynamics, especially for brief or multi-stop trips. Third, although GPS-based location data from a large sample over an extended period offers significant analytical power, the dataset lacks contextual information about users’ trip purposes. Knowing why individuals visit certain places---e.g., for work, shopping, or leisure---would enhance the interpretability of observed patterns, especially if such intent could be self-reported via mobile applications. Finally, the dataset does not include users’ sociodemographic attributes, land use types, or road network characteristics. Access to such information, as available in some unanonymized or linked datasets, could allow researchers to examine behavioral differences across population subgroups or assess the influence of the built environment on mobility decisions.

\section{Methods}\label{sec:methods}

This section details the major methods and data used in our study. 

\subsection{Data Description and Preprocessing}

Details about the dataset and preprocessing steps are provided in SI. In brief, this study uses the YJMob100K dataset, which contains anonymized mobility trajectories of 25,000 individuals in a Japanese metropolitan area over 75 days in 2023, including 15 days during an emergency period. Location data are discretized into 500-meter grid cells within a 100 km × 100 km area and recorded at 30-minute intervals. The study focuses on comparing behavior across 15 usual days and 15 emergency days to examine shifts in mobility under crisis conditions.

\subsection{Mobility Indicators and Computation}

To analyze variations in human movement across different time periods, we focus on the radius of gyration ($r_g$), which measures the typical distance traveled by a user from the center of mass of their trajectory. The radius of gyration provides a comprehensive view of an individual’s typical movement pattern and is particularly useful in characterizing mobility behavior during both normal and emergency periods.

The total radius of gyration is calculated as follows \cite{gonzalez2008understanding,pappalardo2015returners}:

\begin{equation}
r_g = \sqrt{\frac{1}{N} \sum_{i=1}^L n_i \left[ (x_i - x_{cm})^2 + (y_i - y_{cm})^2 \right]}
\end{equation}

where \(L\) denotes the total number of unique visited locations, \((x_i, y_i)\) are the coordinates of location \(i\), and \(n_i\) is the number of times location \(i\) was visited. The center of mass of the trajectory is represented by \((x_{cm}, y_{cm})\), and \(N = \sum_{i=1}^{L} n_i\) is the total number of visits. This metric captures the typical spatial extent of an individual's movements.

To assess how an individual's most frequented locations contribute to their overall mobility range, the \(k\)-radius of gyration is computed \cite{pappalardo2015returners}:

\begin{equation}
r_g^{(k)} = \sqrt{\frac{1}{N_k} \sum_{i=1}^k n_i \left[ (x_i - x_{cm}^{(k)})^2 + (y_i - y_{cm}^{(k)})^2 \right]}
\end{equation}

where \(k\) denotes the number of most visited locations, \(n_i\) is the number of visits to location \(i\), and \((x_{cm}^{(k)}, y_{cm}^{(k)})\) is the center of mass of the top \(k\) locations. \(N_k\)  is the sum of the weights assigned to the k-th most frequented locations. This metric captures the mobility range based on the top \(k\) locations; when \(r_g^{(k)} \approx r_g\), the individual's mobility is largely defined by  their top \(k\) most frequented places.

To classify individuals into \textit{returners} and \textit{explorers}, we employ the \(k\)-radius of gyration in conjunction with a bisector-based threshold. An individual is classified as a \(k\)-returner if \(r_g^{(k)} < (r_g/2)\), suggesting that their mobility is concentrated around their top \(k\) most visited locations. Conversely, individuals with \(r_g^{(k)} > (r_g/2)\) are classified as \(k\)-explorers, indicating a more spatially dispersed travel pattern beyond their primary destinations \cite{pappalardo2015returners}.

\subsubsection{Distribution Fitting}

Three probability distribution functions are employed to characterize distributions of the radius of gyration: the exponential distribution, the lognormal distribution, and the truncated power-law distribution. These functions are defined as follows:

The probability density function for the exponential distribution takes the form:
\begin{equation}
    P_1(x) = \lambda e^{-\lambda x}, \quad x \geq 0
\end{equation}

where $\lambda > 0$ is the rate parameter controlling the decay.

The lognormal distribution has the form:
\begin{equation}
    P_2(x) = \frac{1}{x \sigma \sqrt{2\pi}} \exp\left(-\frac{(\ln x - \mu)^2}{2\sigma^2}\right), \quad x > 0
\end{equation}

where $\mu$ is the mean of the logarithm of the variable, and $\sigma$ is the standard deviation of the logarithm of the variable.

The truncated power-law distribution is defined as:
\begin{equation}
    P_3(x) = \frac{\lambda^{1-\alpha}}{\Gamma(1-\alpha, \lambda x_{\text{min}})} x^{-\alpha} e^{-\lambda x}, \quad x \geq x_{\text{min}}
\end{equation}

where $\alpha$ is the power-law scaling exponent, $\lambda$ is the exponential cutoff parameter, $x_{\text{min}}$ is the lower bound of the distribution, and $\Gamma(1-\alpha, \lambda x_{\text{min}})$ is the upper incomplete Gamma function, ensuring proper normalization.

\subsubsection{Entropy-Based Analysis of Mobility Regularity}

In addition to distributional analysis, we assess the regularity and complexity of individual movement patterns using the real entropy of their trajectories. Entropy reflects the frequency of visits, the sequence of movements, and the duration spent at various locations \cite{qin2012patterns,song2010limits}. The real entropy $E(u)$ for an individual $u$ is calculated as \cite{he2024returners,qin2012patterns,song2010limits}:

\begin{equation}
    E(u) = -\sum_{T_u'} P(T_u') \log_2 P(T_u')
\end{equation}

where $P(T_u')$ denotes the probability of observing a specific time-ordered sequence of locations $T_u'$ within the trajectory of individual $u$.
Here higher entropy values indicate more unpredictable and diverse mobility patterns. Lower entropy values suggest routine and repetitive behavior.


\backmatter

\bmhead{Data Availability}
 
The YJMob100K dataset used in this study is publicly available at Zenodo \cite{yabe2024yjmob100k}.

\bmhead{Code Availability}
Full reproducibility is supported via our open-source codebase on \textcolor{blue}{\href{https://github.com/wissamkontar}{GitHub}}

\bmhead{Author Contribution}
The authors confirm contribution to the paper as follows: study conception and design: Armantalab (lead) and Kontar; data collection: N/A; data processing:  Armantalab (lead) and Kontar; analysis and interpretations of results:  Armantalab (lead), Hawkins, and Kontar; drafting and editing:  Armantalab (lead), Hawkins, and Kontar. All authors reviewed the results and approved the final version of the manuscript.

\bmhead{Competing Interests}
The authors declare no competing interests.

\bmhead{Supplementary information}

\bibliography{references}
\end{document}


\captionsetup[figure]{ labelsep=colon, 
  labelfont=bf, 
  textfont=normalfont, 
  format=plain,
  name=Supplementary Figure,
  labelformat=default
}

\captionsetup[table]{
  labelfont=bf,
  textfont=normalfont,
  format=plain,
  labelsep=colon,
  name=Supplementary Table
}

\title[Mobility Behavior in Crisis: Returners-Explorers Over Long Emergency Periods]{\textbf{Supplementary Information} for Mobility Behavior Evolution During Extended Emergencies: Returners, Explorers, and the 15-Minute City}

\maketitle

\section*{2 Results}

\subsection*{2.1 Mobility Dynamics: Normal and Emergency}

Supplementary Figure 1 presents the distribution fitting results for the radius of gyration ($r_g$) during both normal and emergency periods, using three candidate distributions: truncated power law, exponential, and lognormal. The maximum likelihood estimates for each fit are summarized in Supplementary Table 1. As shown in the table, the lognormal distribution provides the best fit across both periods. During the emergency period, the estimated lognormal parameters exhibit a slight decrease in $\mu$ and an increase in $\sigma$, indicating a broader spread and subtle changes in mobility behavior compared to the normal period. While earlier studies have often modeled displacement distributions using power-law or truncated power-law distributions \cite{brockmann2006scaling,gonzalez2008understanding}, other research has shown that exponential and lognormal distributions may better capture human mobility patterns \cite{liang2012scaling,alessandretti2020scales}, which is consistent with our findings.

\begin{figure}[!ht]
    \centering
    \includegraphics[width=0.9\linewidth]{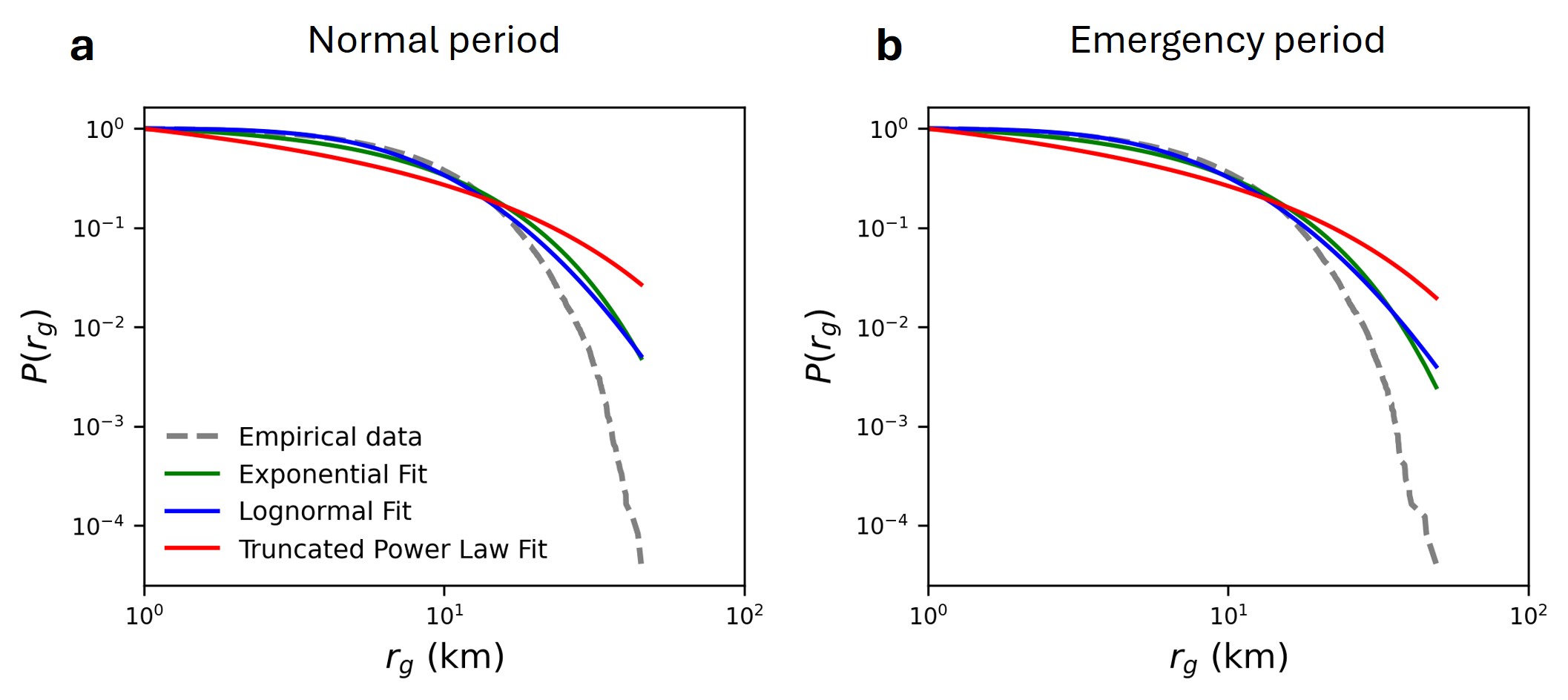}
    \caption{Empirical data of radius of gyration ($r_g$) during normal and emergency periods with fitted candidate distributions: truncated power law, exponential, and lognormal.}
    \label{fig:distributions_rg_plot}
\end{figure}

\begin{table}[!ht]
    \caption{Maximum likelihood estimation results. Positive R values indicate a better fit for the lognormal distribution; negative values favor the alternative.}\label{tab:fit_results}
    \centering
    \footnotesize
    \begin{tabular}{l c c c}
        \hline
        \textbf{Study Period} & \makecell{Lognormal Parameters\\($\mu$, $\sigma$)}& \makecell{Lognormal vs\\Truncated Power Law (R)} & \makecell{Lognormal vs\\Exponential (R)} \\
        \hline
        Normal Period & 2.01, 0.70 & 7423 & 1193 \\
        \hline
        Emergency Period & 1.96, 0.72  & 6428 & 720 \\
        \hline
    \end{tabular}
\end{table}

To examine the consistency of the lognormal fit and its parameters over time, a day-by-day analysis was conducted, and the results are presented in Supplementary Table 2 for each of the 15 days in both the normal and emergency periods. In general, the lognormal distribution provides the best fit for most days. However, the trend of lower $\mu$ and higher $\sigma$ observed in the emergency period during the full-period analysis does not consistently apply on a day-to-day basis. The trend varies across the 15-day period, which contrasts with the findings of a previous study where the trend remained stable during the hurricane event \cite{he2024returners}.

\begin{table}[!ht]
    \caption{Maximum likelihood estimation results for each individual day. Positive R values indicate a better fit for the lognormal distribution; negative values favor the alternative.}\label{tab:fit_results_all_days}
    \centering
    \footnotesize
    \begin{tabular}{l c c c}
        \hline
        \textbf{Day} & \makecell{Lognormal Parameters\\($\mu$, $\sigma$)} & \makecell{Lognormal vs\\Truncated Power Law (R)} & \makecell{Lognormal vs\\Exponential (R)} \\
        \hline
        \textbf{Normal Period} & & & \\
        1 & -56.21, 2.26  & -0.01 & 0.15 \\
        2 & 2.51, 0.46  & 2.58 & -5.40 \\
        3 & 2.90, 0.33  & 8.62 & 4.53 \\
        4 & 3.17, 0.25  & 0.46 & 0.34 \\
        5 & -86.56, 2.35  & 1.12 & 1.40 \\
        6 & 2.86, 0.32  & 0.13 & 0.03 \\
        7 & 3.11, 0.26  & 0.13 & 0.10 \\
        8 & 3.16, 0.25  & 0.77 & 0.57 \\
        9 & 2.69, 0.39  & 6.14 & -0.30 \\
        10 & 2.63, 0.42  & 5.06 & -3.12 \\
        11 & 2.73, 0.38  & 5.82 & 0.08 \\
        12 & 3.11, 0.25  & 2.27 & 1.62 \\
        13 & 3.25, 0.22 & 2.12 & 1.72 \\
        14 & 3.18, 0.27  & 2.99 & 2.10 \\
        15 & 2.96, 0.29  & 0.08 & -0.05 \\
        \hline
        \textbf{Emergency Period} & & & \\
        1 & 2.66, 0.41  & 74.13 & -0.69 \\
        2 & 3.14, 0.23  & 4.86 & 0.60 \\
        3 & 2.94, 0.31  & 0.45 & 4.97 \\
        4 & 3.29, 0.19  & 1.35 & 0.63 \\
        5 & 2.92, 0.32  & 42.29 & 5.15 \\
        6 & 3.11, 0.24  & 4.09 & 0.14 \\
        7 & 2.93, 0.28  & 1.13 & -0.02 \\
        8 & 3.02, 0.29  & 19.10 & 2.13 \\
        9 & 2.80, 0.36  & 1.36 & 0.04 \\
        10 & 2.93, 0.27 & 1.75 & 0.01 \\
        11 & 1.30, 0.45  & 0.01 & 0.05 \\
        12 & 2.63, 0.42  & 112.10 & -2.24 \\
        13 & 2.56, 0.34  & 0.21 & 0.01 \\
        14 & 3.25, 0.22  & 2.86 & 0.65 \\
        15 & 2.67, 0.33 & 0.58 & -0.03 \\
        \hline
    \end{tabular}
\end{table}

\subsection*{2.2 Classifying Individuals as Returners or Explorers}

This classification between two mobility patterns—returners and explorers—is based on the $k$-radius of gyration ($r_g^{(k)}$), which quantifies the spatial dispersion of visits to an individual's top $k$ frequently visited locations. For example, $k = 2$ primarily reflects visits to two essential locations, such as home and workplace, which dominate most individuals' routines. By expanding the analysis to $k = 3$, 4, and 5, we capture broader mobility patterns by incorporating additional frequently visited locations. In the following analysis, we examine distribution fits, density patterns, and parameter estimates based on $r_g^{(k)}$ using the top 2, 3, 4, and 5 most frequently visited locations. Since results for $k > 5$ show similar trends to those at $k = 5$, they are not presented.

The distribution fit results are presented in Supplementary Figure 2, with detailed fitting parameters provided in Supplementary Table 3. As $k$ increases, the fitted curves for $r_g^{(k)}$ become increasingly similar to those of the total $r_g$ in both the normal and emergency periods. This indicates that the mobility captured by $r_g^{(k)}$ more closely approximates an individual's full mobility range as additional frequently visited locations are included. Supplementary Table 3 further supports this observation, showing that the estimated values of $\mu$ and $\sigma$ for $r_g^{(k)}$ progressively converge toward those of the total $r_g$ as $k$ increases in both periods.

\begin{figure}[!ht]
    \centering
    \includegraphics[width=1\linewidth]{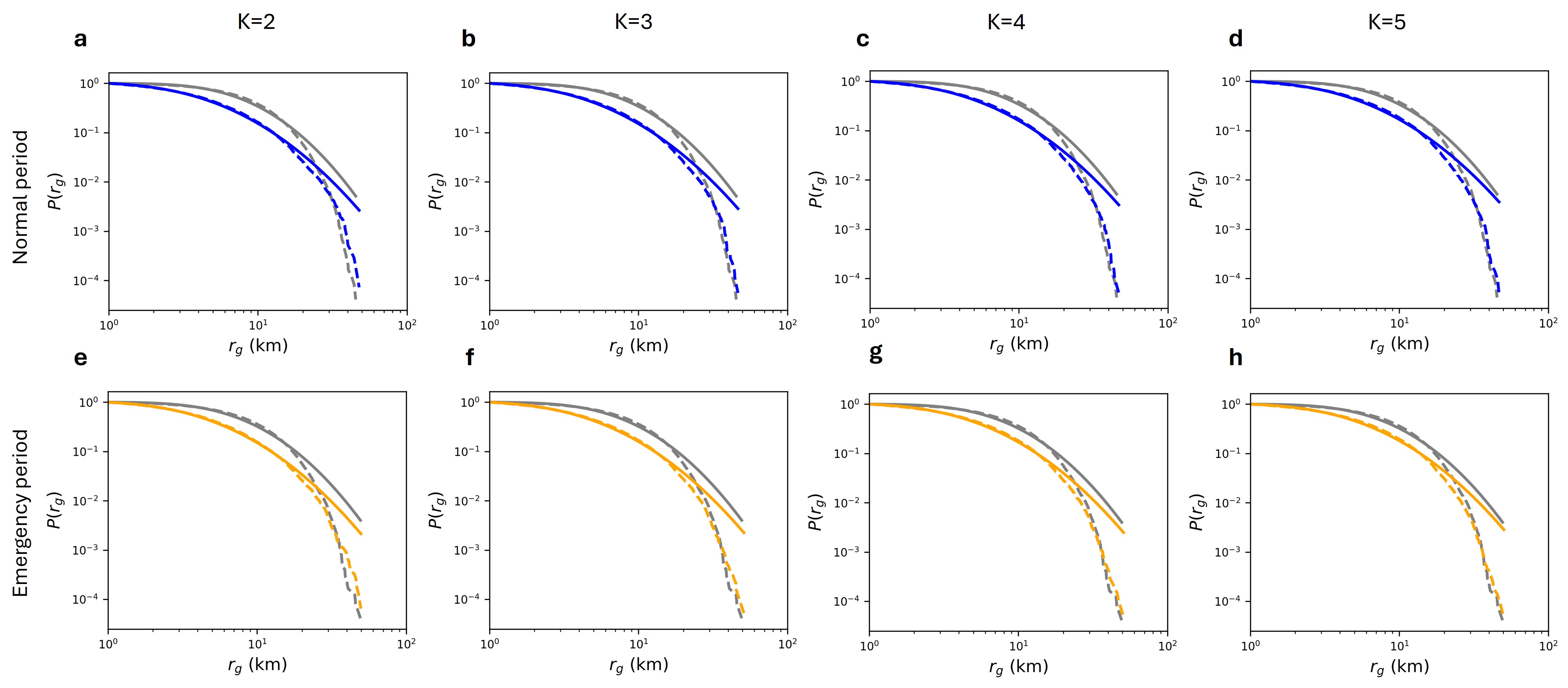}
    \caption{Empirical data (dashed lines) and fitted lognormal curves (solid lines) of total radius of gyration ($r_g$, shown in grey) and $k$-radius of gyration ($r_g^{(k)}$, shown in blue and orange) for different values of $k$ during normal and emergency periods.}
    \label{fig:distributions_plot}
\end{figure}

\begin{table}[!ht]
    \caption{Parameter estimates across top-$k$ radius of gyration values ($k=2,3,4,5$)}\label{tab:fit_results_k}
    \centering
    \footnotesize
    \begin{tabular}{l c c c c}
        \hline
        & \multicolumn{2}{c}{Normal Period} & \multicolumn{2}{c}{Emergency Period} \\
        \hline
        & $\mu$ & $\sigma$ & $\mu$ & $\sigma$ \\
        \hline
        r$_2$ & 1.34 & 0.89 & 1.34 & 0.88 \\
        r$_3$ & 1.33 & 0.90 & 1.37& 0.89 \\
        r$_4$ & 1.36 & 0.89 & 1.42 & 0.88 \\
        r$_5$ & 1.40 & 0.90 & 1.47 & 0.87 \\
        r$_g$ & 2.01 & 0.70 & 1.96 & 0.72 \\
        \hline
    \end{tabular}
\end{table}

In Supplementary Figure 4, a density analysis comparing $r_g^{(k)}$ and $r_g$ is presented. This analysis highlights significant differences between the two mobility groups across varying values of $k$. In both time periods, data points primarily cluster around the diagonal and horizontal axes, suggesting that for one group, travel distances are closely related to their top $k$ locations, indicative of k-returner behavior. Conversely, the other group, known as k-explorers, displays movement patterns that are less associated with their top $k$ locations. As $k$ increases, the clustering becomes more diagonal, signaling a transition from k-explorer to k-returner behavior.

\begin{figure}[!ht]
    \centering
    \includegraphics[width=1\linewidth]{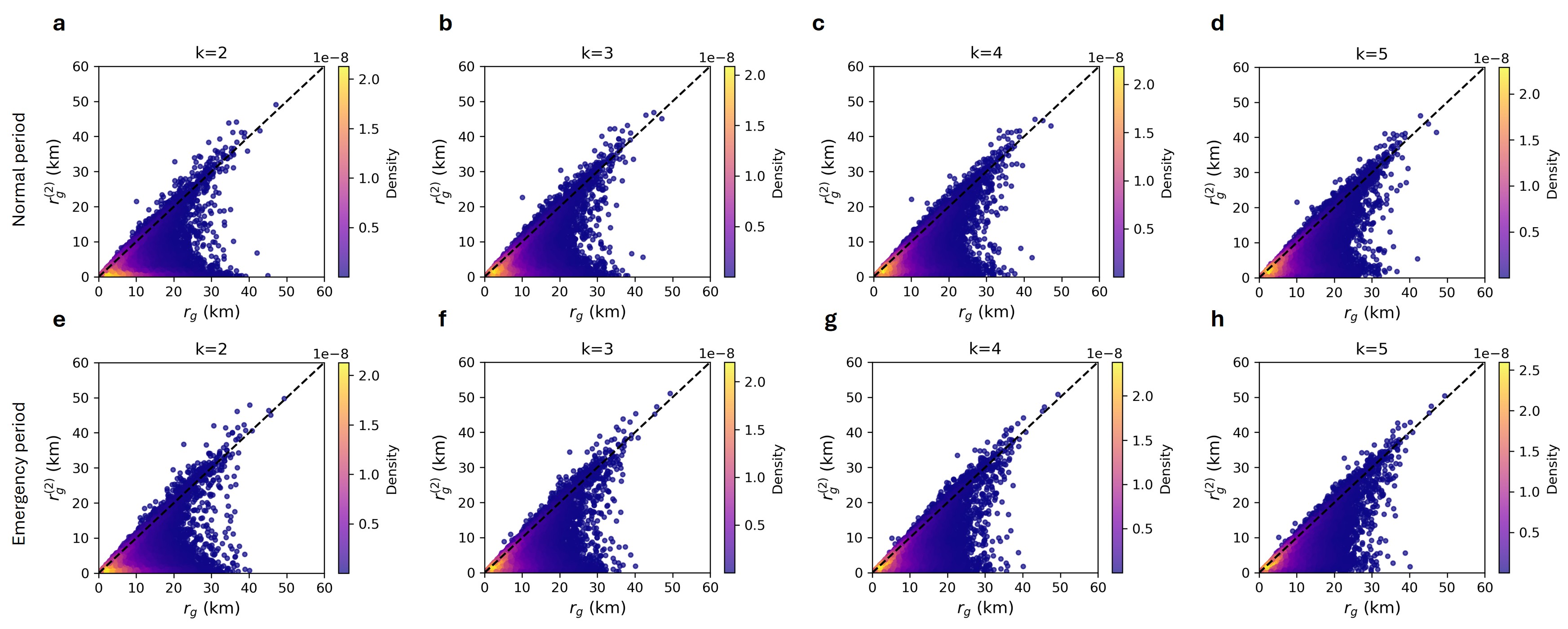}
    \caption{Density scatter plots illustrating the relationship between the total radius of gyration ($r_g$) and the top-$k$ radius of gyration ($r_g^{(k)}$).}
    \label{fig:scatter}
\end{figure}

Three distinct 14-day windows—other than the one used for the main analysis—were selected from the normal period to test whether our findings are sensitive to the choice of the specific 14-day segment. For each of these segments, we conducted the analysis for both the first 7 days and the full 14-day duration. As shown in Supplementary Figure 4, all segments consistently exhibit the same pattern observed in the main text: at 14 days, the crossover point where returners exceed explorers occurs at $k = 4$. This confirms that our conclusions are robust and not dependent on the specific 14-day window selected.

\begin{figure}[!ht]
    \centering
    \includegraphics[width=1\linewidth]{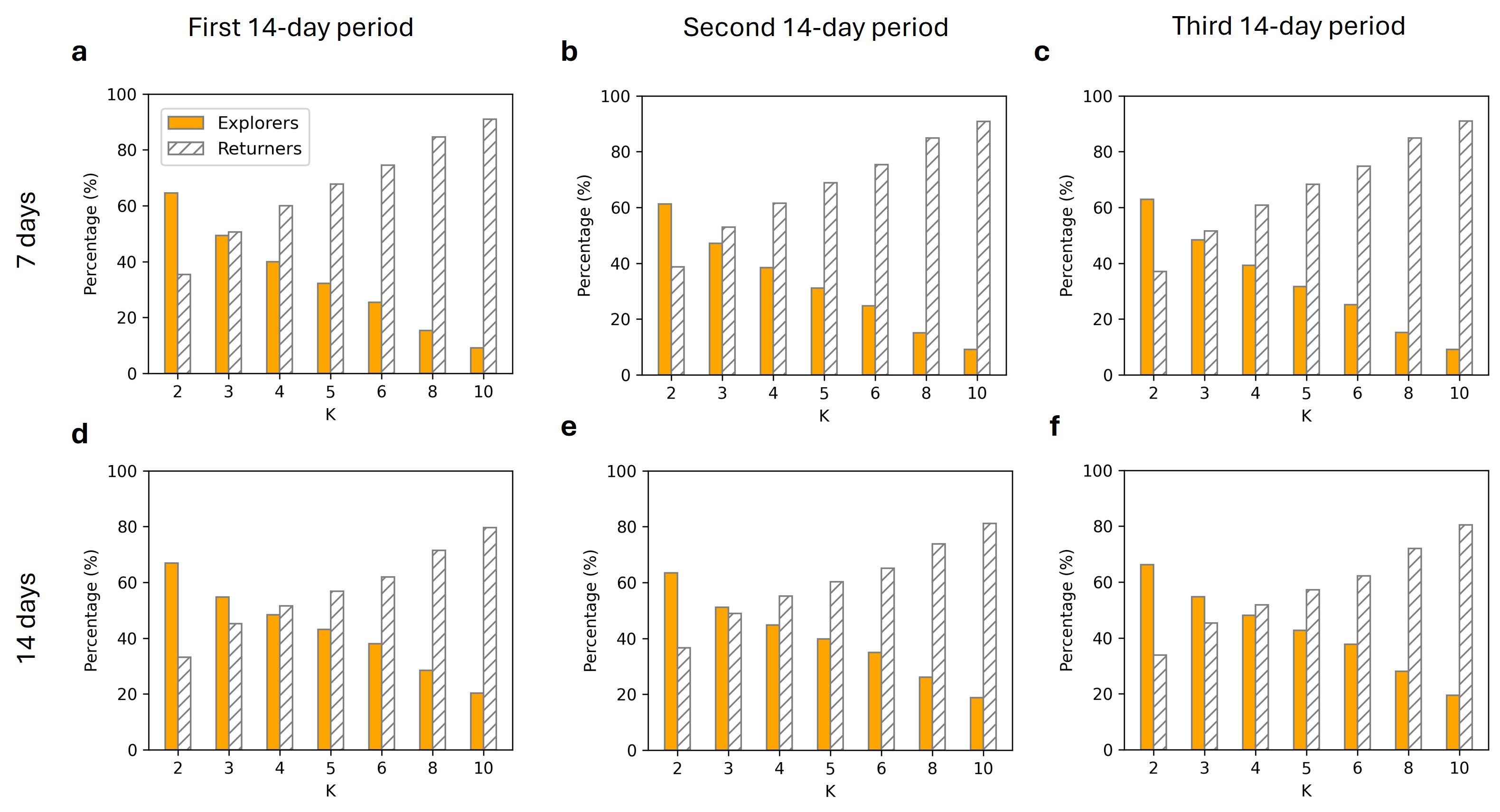}
    \caption{Percentage of $k$-returners and $k$-explorers over the first 7 days and full 14 days for three distinct 14-day segments of the normal period: days 1–14, days 15–28, and days 29–42.}
    \label{fig:sensitivity_1}
\end{figure}

In addition, we explored how extending the observation period in the normal phase impacts the returner--explorer relationship. The results, shown in Supplementary Figure 5, reveal that as the observation period lengthens, the crossover point shifts to higher $k$ values: $k = 4$ for 2 weeks, $k = 5$ for 4 weeks, and $k = 6$ for both 6 and 8 weeks. 

\begin{figure}[!ht]
    \centering
    \includegraphics[width=0.7\linewidth]{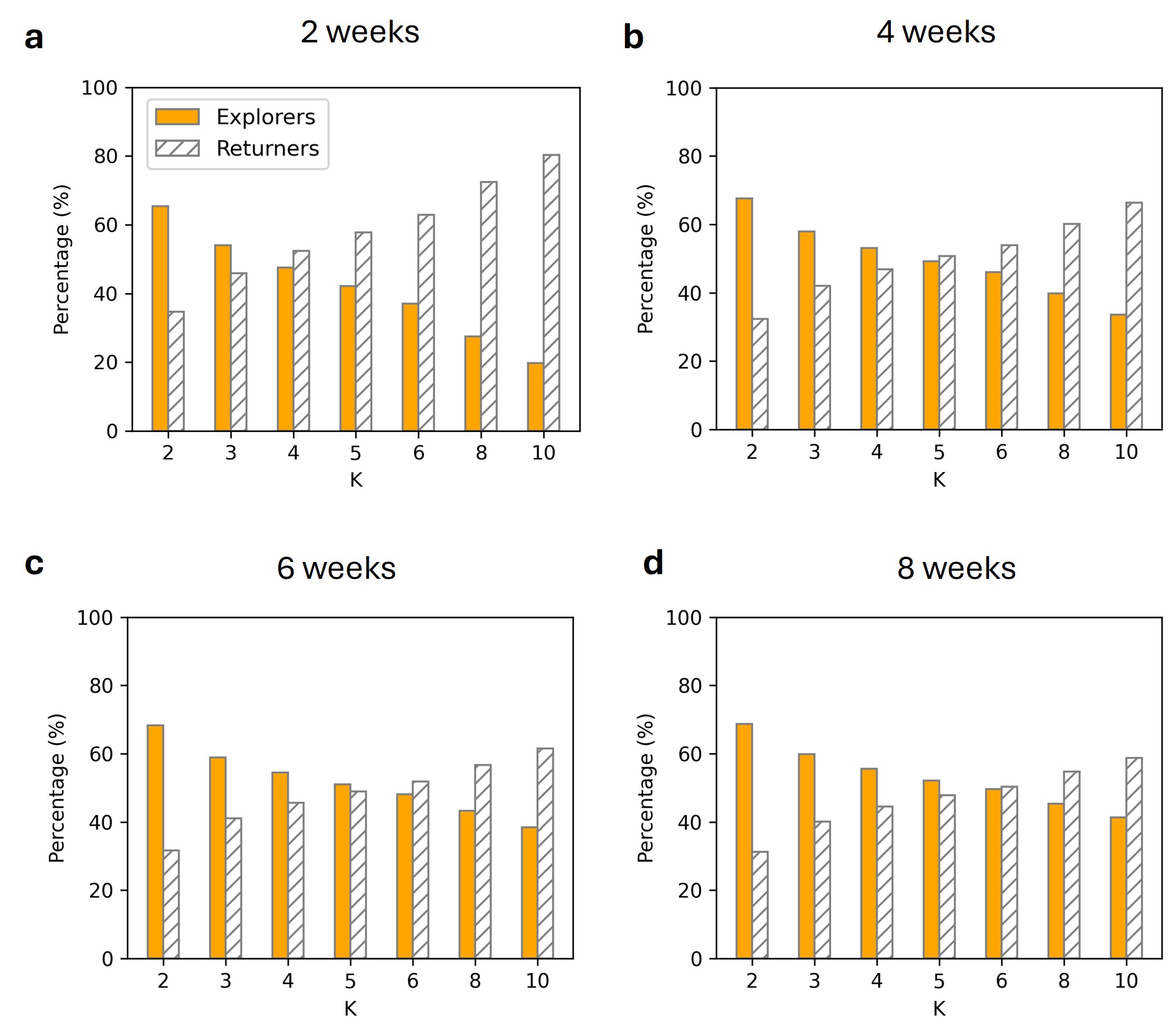}
    \caption{Percentage of $k$-returners and $k$-explorers during extended observation windows in the normal period: 2 weeks (days 43–56), 4 weeks (days 29–56), 6 weeks (days 15–56), and 8 weeks (days 1–56).}
    \label{fig:sensitivity_2}
\end{figure}

\subsubsection*{2.3.3 Behavioral transitions}

Supplementary Figure 6 illustrates the daily maximum distance from home for the four mobility groups during the emergency period. On weekends and holidays (e.g., days 6, 7, and 8), a notable behavioral shift is observed among the returner–returner group—individuals who remained returners during the emergency. This shift is characterized by a substantial increase in the percentage of short-distance trips and a corresponding decrease in longer maximum distances from home, particularly on Sundays (days 7 and 14). Changes among the other groups are less pronounced.

\begin{figure}[!ht]
    \centering
    \includegraphics[width=1\linewidth]{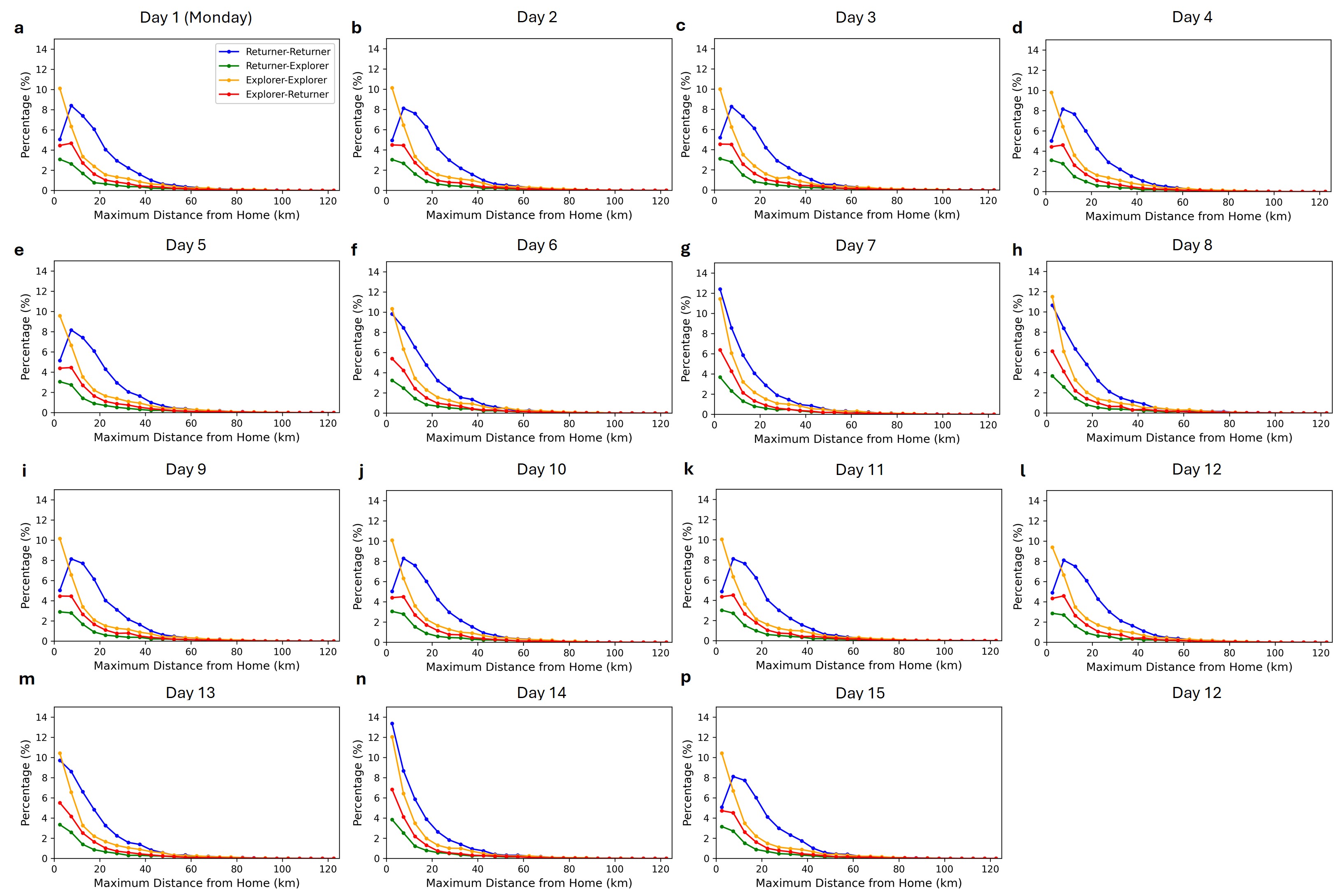}
    \caption{Daily maximum distance from home for four mobility groups during the emergency period.}
    \label{fig:max_dist_4groups}
\end{figure}

Supplementary Figure 7 illustrates the non-home dwelling time for each of the four groups over the emergency period. On weekends and holidays—especially Sundays (e.g., days 7 and 14)—all four groups exhibit a decrease in long non-home dwelling durations and an increase in shorter ones. This shift is most pronounced for the returner–returner group, indicating that individuals across all groups tend to favor short-duration non-home activities on holidays. On day 15, the final day of the emergency period, returner-returner group shows a notable rise in medium-duration non-home activities and a decline in very long ones, indicating behavioral adaptation as the emergency concludes.

\begin{figure}[!ht]
    \centering
    \includegraphics[width=1\linewidth]{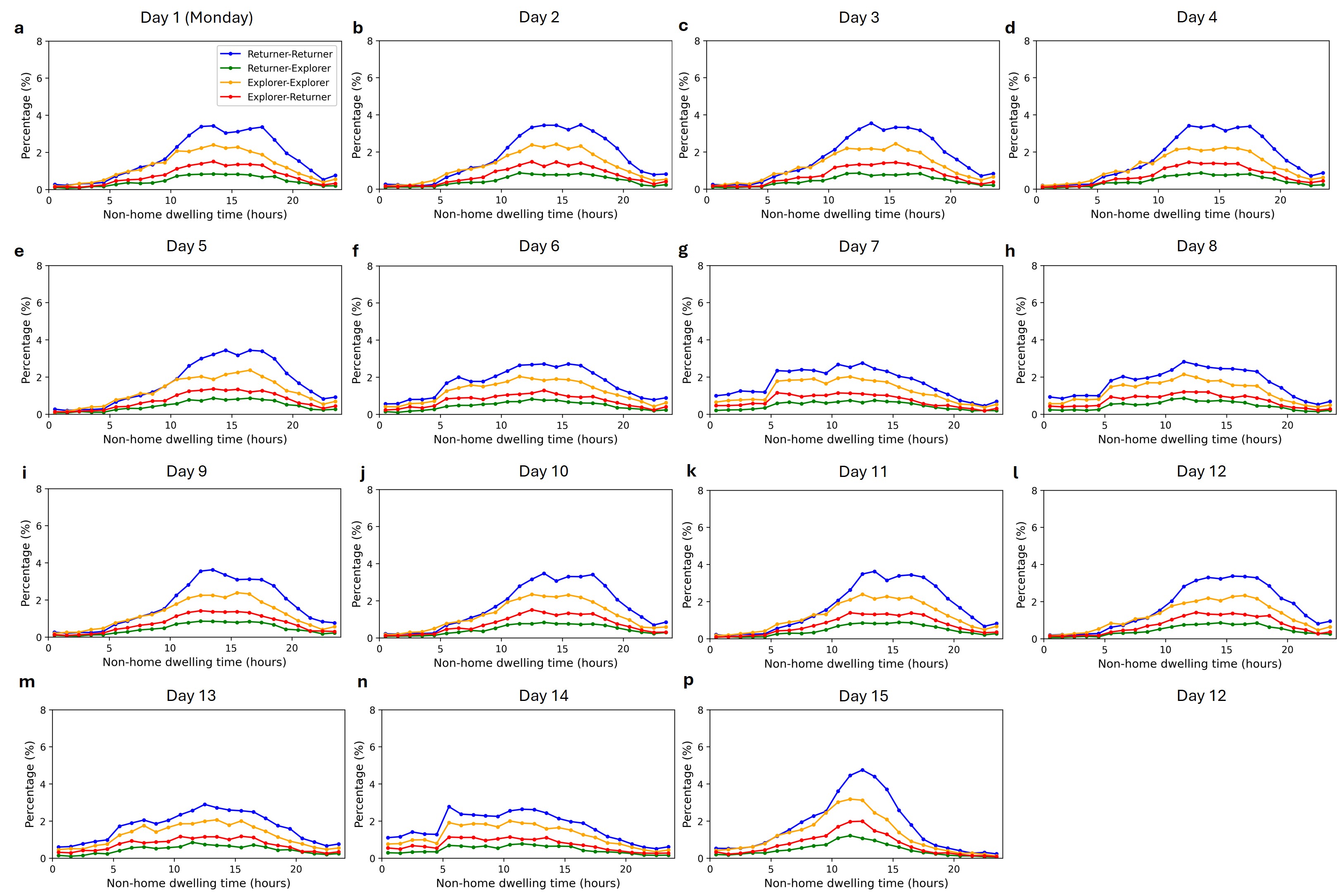}
    \caption{Daily non-home dwelling time for four mobility groups during the emergency period.}
    \label{fig:time_away_4groups}
\end{figure}

\subsubsection*{2.3.4 Entropy}

To further explore the differences between the two groups, this study also examines the per capita real entropy across two distinct periods, as well as how these values change in response to the impacts of the emergency period. The results shown in Supplementary Figure 8 reveal noticeable discrepancies between the groups across the two periods, with Supplementary Table 4 confirming the statistical significance of these differences. Across both periods, explorers consistently exhibit a higher per capita real entropy than returners. The movement patterns of explorers are marked by greater variability and unpredictability, while returners demonstrate more consistent and stable behavior.

\begin{figure}[!ht]
    \centering
    \includegraphics[width=0.9\linewidth]{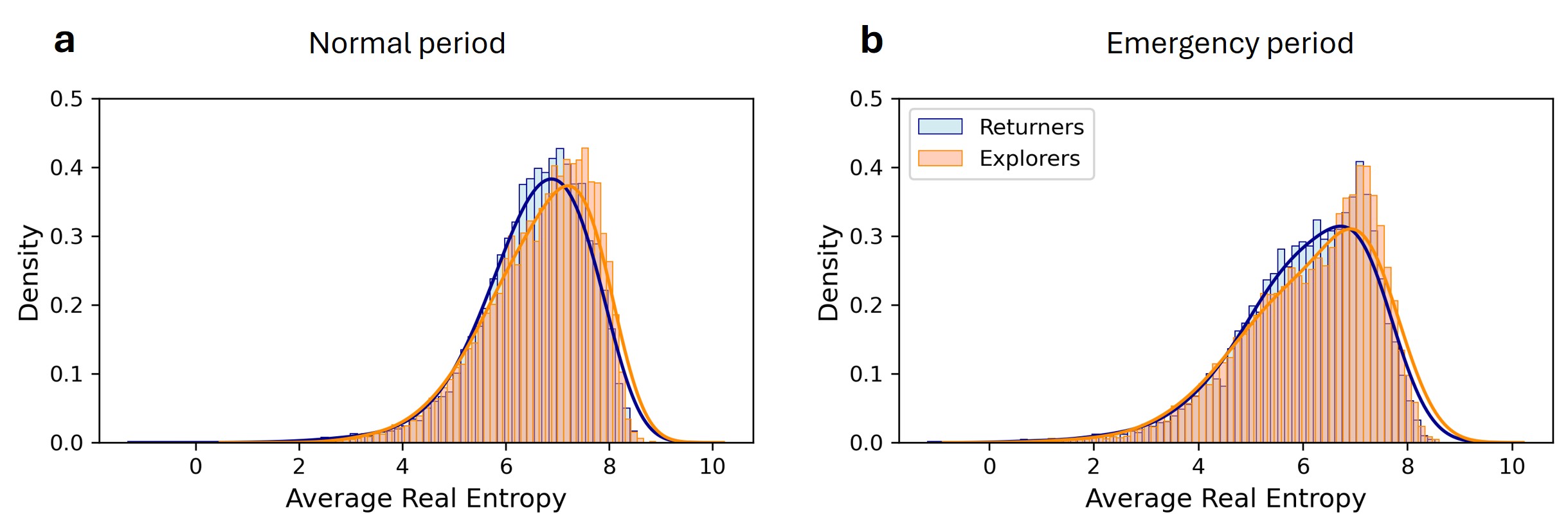}
    \caption{Density distributions of average real entropy for returners and explorers during the normal (a) and emergency (b) periods, shown with kernel density estimate (KDE) curves.}
    \label{fig:entropy}
\end{figure}

\begin{table}[!ht]
    \caption{Kolmogorov-Smirnov and Mann-Whitney U test results: real entropy differences between returners and explorers during normal and emergency periods}\label{tab:significance_test_entropy}
    \centering
    \footnotesize
    \begin{tabular}{l c c c c}
        \hline
        & \multicolumn{2}{c}{Mann-Whitney U test} & \multicolumn{2}{c}{Kolmogorov-Smirnov test} \\\hline
         & P-value & Significance & P-value & Significance \\
        \hline
        \textbf{Emergency} & 0.0 & **& 0.0 & ** \\
        \hline
        \textbf{Normal} & 0.0 & **& 0.0 & ** \\
        \hline
        Note: ** means significant at the 0.01 level.
    \end{tabular}
\end{table}

\subsection*{2.4 Spatiotemporal mobility and 15-minute city}

Supplementary Figure 9 illustrates that, during both the emergency and normal periods, returners exhibit significantly higher numbers of stops per person in the central area of the major city, located around coordinates $x = 68$ and $y = 38$. Overall, in both periods, the number of stops per person for both returners and explorers tends to concentrate more in this central area. However, during the emergency period, the number of stops for both groups increases in areas outside the city center. This suggests that, in the emergency period, mobility becomes more dispersed across other areas, whereas, in the normal period, it is more centralized in the city's core.

\begin{figure}[!ht]
    \centering
    \includegraphics[width=1\linewidth]{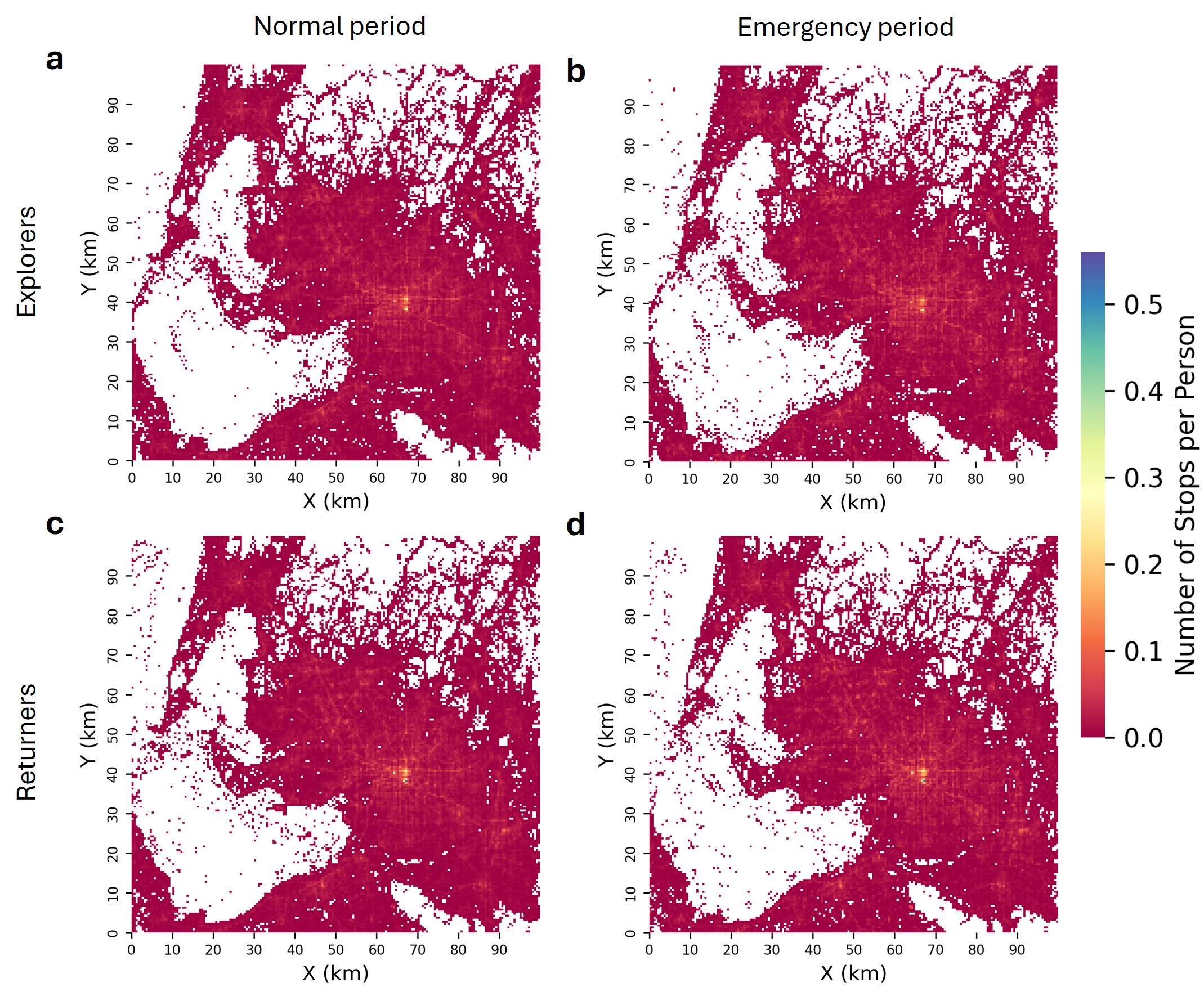}
    \caption{Comparison of spatial distributions of activity stops for explorers (a, b) and returners (c, d) during normal and emergency periods.}
    \label{fig:spatial_stops}
\end{figure}

Supplementary Figure 10 shows that the average stay time at stops is notably higher for returners during both the normal and emergency periods, particularly in the center of the major city around $x = 68$ and $y = 38$. Comparing the normal period to the emergency period, both returners and explorers exhibit a decrease in average stay time in the city center. This decrease is more pronounced for explorers than for returners, indicating a stronger reduction in their stay time at the city center during the emergency phase.

\begin{figure}[!ht]
    \centering
    \includegraphics[width=1\linewidth]{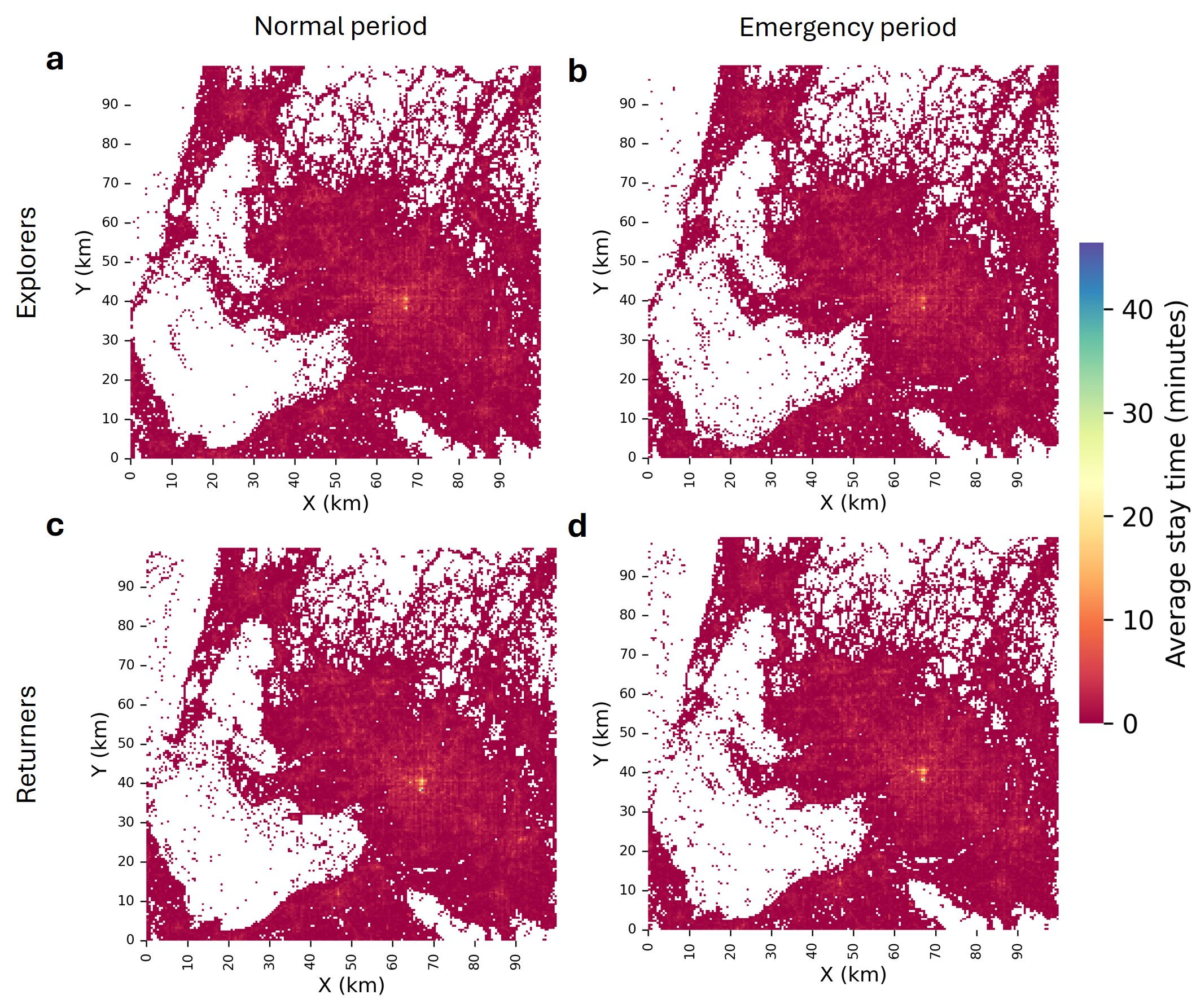}
    \caption{Comparison of average duration of stay at stop locations for explorers (a, b) and returners (c, d) during normal and emergency periods.}
    \label{fig:spatial_times}
\end{figure}

\paragraph{Details on the definition of nearby neighborhoods} Our data comes from a 100 km \(\times\) 100 km metropolitan region, divided into 500 m \(\times\) 500 m grid cells. We define nearby neighborhoods as cells with a maximum difference of 2 units in \(x\) and \(y\) coordinates, representing areas likely within a walkable distance. This spatial definition approximately aligns with the concept of the “15-minute city” \cite{abbiasov202415}, where most daily needs can be met within a 15-minute walk from home.

\section*{4 Methods}
\subsection*{4.1 Data Description and Preprocessing}

This study utilizes the YJMob100K dataset, an open-source and anonymized collection of human mobility trajectories provided by Yahoo Japan Corporation (now LY Corporation) \cite{yabe2024yjmob100k}. The dataset captures the movement patterns of 25,000 individuals over a 75-day period in 2023, in a densely populated and undisclosed metropolitan area in Japan. Of these 75 days, 60 represent typical, business-as-usual behavior, while the remaining 15 correspond to an emergency period characterized by unusual mobility patterns.

To contextualize this division, supplementary Figure 1 illustrates the number of stops made by all individuals each day over the 75-day period. A consistent pattern of reduced stops is observed every seven days, corresponding to weekends, primarily Saturdays and Sundays, across both normal and emergency phases. In addition to weekends, a few sharp declines in stop counts exist on weekdays, which are likely holidays exhibiting similar activity reductions. These recurring patterns help distinguish between typical weekdays, weekends, and holidays. For comparative analysis, days 43 to 57 were selected as the representative normal period, while days 60 to 74 define the emergency period. Both periods were chosen to start on a Monday and include a holiday on the following Monday (i.e., day 8), ensuring comparable weekly structures. 

\begin{figure}[!ht]
    \centering
    \includegraphics[width=0.9\linewidth]{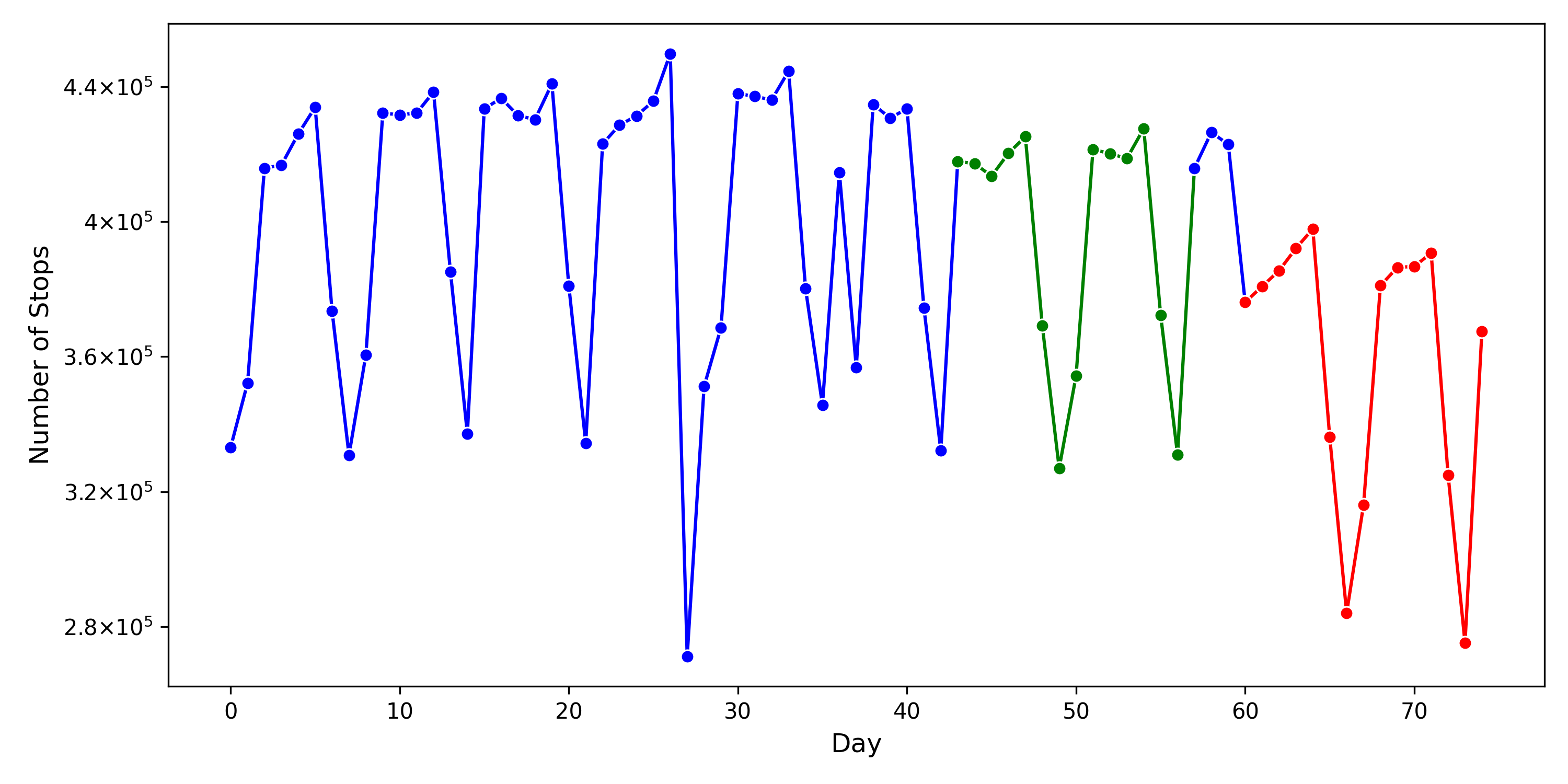}
    \caption{Daily number of stops made by all individuals over the 75-day period. Green highlights the selected 15-day normal period, and red highlights the 15-day emergency period.}
    \label{fig:activity}
\end{figure}

Each human trajectory record includes a user ID, day, time, and the $(x, y)$ grid cell indicating the user’s location at that time. The spatial grid consists of $x \in [1, 200]$ and $y \in [1, 200]$, where each cell represents a 500-meter by 500-meter area, covering a total region of 100 km $\times$ 100 km. Time is discretized into 30-minute intervals across 75 days (day 0 to day 74). The dataset maintains a sampling rate of approximately 5\% of the population within the study area, ensuring spatial representativeness \cite{yabe2024yjmob100k}. Due to anonymization, actual geographic coordinates and calendar dates are not disclosed. 

To enrich the spatial context, we incorporated a supplementary dataset from the same corporation, which provides the number of Points of Interest (POIs) in each grid cell. Additionally, each individual’s home location was inferred using their movement patterns during the 60 usual days. Following the dataset manual \cite{yabe2024yjmob100k}, the home location was defined as the grid cell with the highest frequency of visits between 8:00 PM and 8:00 AM over the normal period.

\clearpage
\newpage
\bibliography{references}